\documentclass[nofootinbib,preprint,preprintnumbers,a4paper,11pt]{revtex4}
\usepackage{epsfig}
\usepackage{footnote}
\usepackage{ulem}
\usepackage{color}
\usepackage{array}
\usepackage{amssymb}
\usepackage{amsmath}
\usepackage{graphicx}
\usepackage{subfigure}
\usepackage{longtable}
\usepackage{verbatim}
\usepackage{amsfonts}
\usepackage{hyperref}
\usepackage{multirow}
\usepackage{cancel}

\begin{document}

\title{From topological amplitude to rescattering dynamics}

\author{Di Wang$^{1}$}\email{wangdi@hunnu.edu.cn}

\address{%
$^1$Department of Physics, Hunan Normal University, and Key Laboratory of Low-Dimensional Quantum Structures and Quantum Control of Ministry of Education, Changsha 410081, China
}

\begin{abstract}
We proposed a theoretical framework to correlate the topological diagram at quark level and rescattering dynamics at hadron level.
In the framework, both the hadronic triangle diagram, and the topological-scattering diagram, which is the intermediate structure between topological diagram and triangle diagram, are expressed in the tensor form.
The completeness of topological-scattering diagram is confirmed by the quark substructure of meson-meson scattering.
The coefficient of each triangle diagram can be derived from the topological-scattering diagram and the total rescattering amplitudes are consistent with the ones derived from the chiral Lagrangian.
If only the short-distance $T$ diagram is considered as the weak vertex in triangle diagram, the rescattering contributions in the $C$, $E$ and $P$ diagrams have definite proportional relation of $L(C):L(E):L(P)=-2:1:1$ under the $SU(3)_F$ symmetry, and the rescattering contributions in the $T$ and $A$ diagrams only arise from the $SU(3)_F$ breaking effects.
Taking $D\to K\pi$ and $D\to \pi\pi$ modes as examples, we present our framework in detail.
We find the Isospin relations in these decays are still valid in terms of triangle diagrams.
Besides, the conclusions in the $D$ meson decays under the $SU(3)_F$ symmetry can be generalized to the $B$ meson decays under the $SU(4)_F$ symmetry.
\end{abstract}

\maketitle

\newpage
\section{Introduction}\label{intro}

Heavy meson/baryon non-leptonic decay provides an ideal platform to test the Standard Model (SM) and search for new physics.
Many data have been collected by experiments in the last few decades \cite{ParticleDataGroup:2020ssz}.
In theory, the non-perturbative QCD dynamics is still a challenge, especially in charmed hadron decay.
The QCD-inspired approaches, such as QCD factorization (QCDF) \cite{Beneke:1999br,Beneke:2000ry,Beneke:2003zv,Beneke:2001ev}, perturbative QCD approach (PQCD) \cite{Keum:2000ph,Keum:2000wi,Lu:2000em,Lu:2000hj}, and soft-collinear effective theory (SCET) \cite{Bauer:2001cu,Bauer:2001yt}, do not work well in charm sector because of the large expansion parameters $\alpha_s(m_c)$ and $\Lambda_{\rm QCD}/m_c$.
An alternative way to investigate the charmed hadron decay is the flavor symmetry analysis.
The topological diagram amplitude (TDA) approach \cite{Rizzo:1980yh,Zeppenfeld:1980ex,Chau:1982da,Chau:1986du,Chau:1987tk}, in which the topological diagrams are classified according to the topologies in the flavor flow of weak decay diagrams, is very popular.
Topological diagram is intuitive to the internal dynamics of hadron decays, providing a framework in which we cannot only do a model-dependent data analysis \cite{Cheng:2012xb,Cheng:2012wr,Muller:2015lua,Cheng:2016ejf,Cheng:2014rfa,Cheng:2010ry} but also perform a theoretical model calculation \cite{Zhou:2021yys,Wang:2017hxe,Zhou:2016jkv,Zhou:2015jba,Li:2012cfa,Qin:2013tje,Wang:2017ksn,Jiang:2017zwr,Yu:2017oky}.
In Refs.~\cite{He:2018php,He:2018joe,Wang:2020gmn}, a systematic method to treat the topological amplitudes was proposed.
In the framework, the topologies are expressed in invariant tensors and the universality, completeness and correlation of topology are clarified.

\begin{figure}[htp]
  \centering
  \includegraphics[width=11cm]{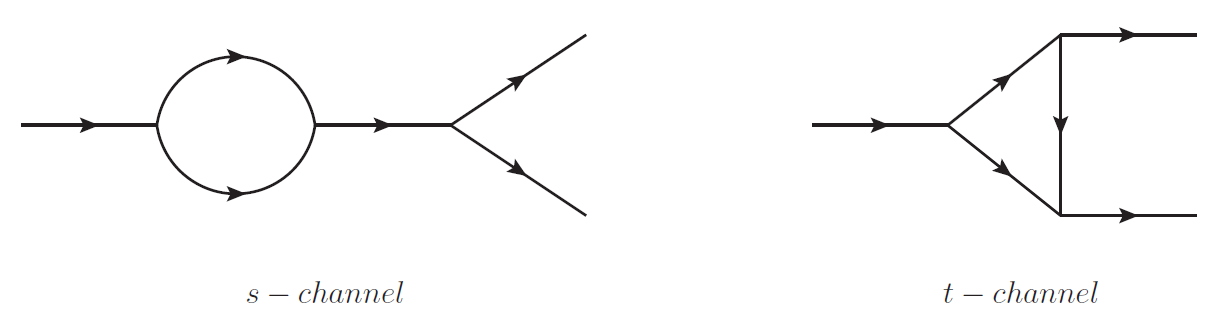}
  \caption{$s$-channel (resonance) and $t$-channel (rescattering) contributions to final-state interaction.}\label{tx}
\end{figure}

Another approach to study the non-perturbative QCD effects is the final-state interaction (FSI).
In this framework, the non-perturbative QCD effects are modeled as an exchange of one particle between two particles generated from the short-distance tree emitted process. There are s-channel and t-channel contributions in the final state interaction, or referred to as resonance and rescattering contributions respectively.
The diagrams of these two nonperturbative effects can be depicted in Fig.~\ref{tx}.
In this work, we focus on the rescattering contribution.
It forms a triangle diagram at hadron level.
The rescattering dynamics was used to estimate the branching fractions of heavy mason and baryon hadron decays \cite{Cheng:2002wu,Han:2021gkl,Locher:1993cc,Li:1996cj,Dai:1999cs,Li:2002pj,Ablikim:2002ep,Cheng:2004ru,Lu:2005mx,Chen:2002jr}.
In 2017, the rescattering dynamics was extended to the doubly charmed baryon decays, helping us observe $\Xi_{cc}^{++}$ for the first time \cite{Yu:2017zst,LHCb:2017iph}.
After that, the branching fractions of other doubly charmed baryon decay channels were estimated in the rescattering dynamics \cite{Jiang:2018oak,Han:2021azw}.

The topological amplitude and rescattering dynamics seems to be two unrelated theories since one is at the quark level and the other is at the hadron level.
In \cite{Cheng:2002wu,Ablikim:2002ep,Cheng:2004ru}, the authors attempted to establish the relation between the topological diagram and triangle diagram via the intermediate structure between them, we called it topological-scattering diagram in this work.
Because of the absence of a systematic method, they did not find the topological-scattering diagrams completely, resulting in some confusions.
For example, the Isospin factor $1/\sqrt{2}$ or $-1/\sqrt{2}$ is added manually in  \cite{Ablikim:2002ep} to satisfy Isospin relation of the $D^+\to \pi^+\pi^0$, $D^0\to \pi^+\pi^-$ and $D^0\to \pi^0\pi^0$ modes without a convictive argument.
And the triangle diagrams extracted by topological-scattering diagrams are in conflict with the ones derived from the chiral Lagrangian \cite{Cheng:2004ru}.
Thereby, a systematic study of the relation between the topological amplitude and the rescattering dynamics is necessary.

Inspired by Refs.~\cite{He:2018php,He:2018joe,Wang:2020gmn}, we find that the topological-scattering diagram can also been expressed in tensor form just like topological diagram.
Meanwhile, the topological-scattering diagram written in tensor form can be understood as triangle diagrams in the rescattering dynamics.
As a result, the algebraic tensor serves as a bridge between the topological diagram at quark level and the triangle diagram in hadron level.
The completeness of the topological-scattering diagram is confirmed by the quark substructure of meson-meson scattering.
The coefficient of each triangle diagram, such as Isospin factor, can be derived from the topological-scattering diagram.
The triangle diagrams derived from the topological diagrams are consistent with the ones derived from the chiral Lagrangian.

Taking $D\to K\pi$ and $D\to \pi\pi$ modes as examples, we present our framework in detail.
We find the triangle diagrams with vertexes of $\omega\pi\pi$, $\rho^0\pi^0\pi^0$ ... cancel each other after summing all triangle diagrams in one decay channel.
The Isospin relations hold in terms of the triangle diagrams.
Under the $SU(3)_F$ symmetry, the rescattering contributions arisen from $T^{SD}$ in the $C$, $E$ and $P$ diagrams have the relation of $L(C):L(E):L(P)=-2:1:1$.
If the $SU(3)_F$ symmetry breaks into the Isospin symmetry, the proportional relation is broken.
The rescattering contributions arisen from $T^{SD}$ in the $T$ and $A$ diagrams only arise from the $SU(3)_F$ breaking effects. And there are no triangle diagram like long-distance contributions in the topologies $ES$, $AS$, $PS$, $PA$ and $SS$.

This paper is organized as follows. In Sec.~\ref{ttor}, we construct the tensor representation of topological-scattering diagram and triangle diagram, and investigate the relation between topological amplitude and rescattering dynamics.
Taking $D\to K\pi$ and $D\to \pi\pi$ decays as examples, we show the reliability of our method and give some conclusions in Sec.~\ref{cdecay}.
Sec.~\ref{sum} is a short summary.
Further discussions in the $\overline B\to D\pi$ and $\overline B\to \pi\pi$ decays can be found in Appendix.~\ref{bdecay}.

\section{From topological diagram to triangle diagram}\label{ttor}

In this section, we first review the basic idea of the topological diagram expressed in the invariant tensor, taking the $D\to PP$ decay as examples.
Then we try to write the topological-scattering diagram and triangle diagram in the tensor form and analyze the relation among topological diagram, topological-scattering diagram and triangle diagram.

\begin{figure}
  \centering
  \includegraphics[width=15cm]{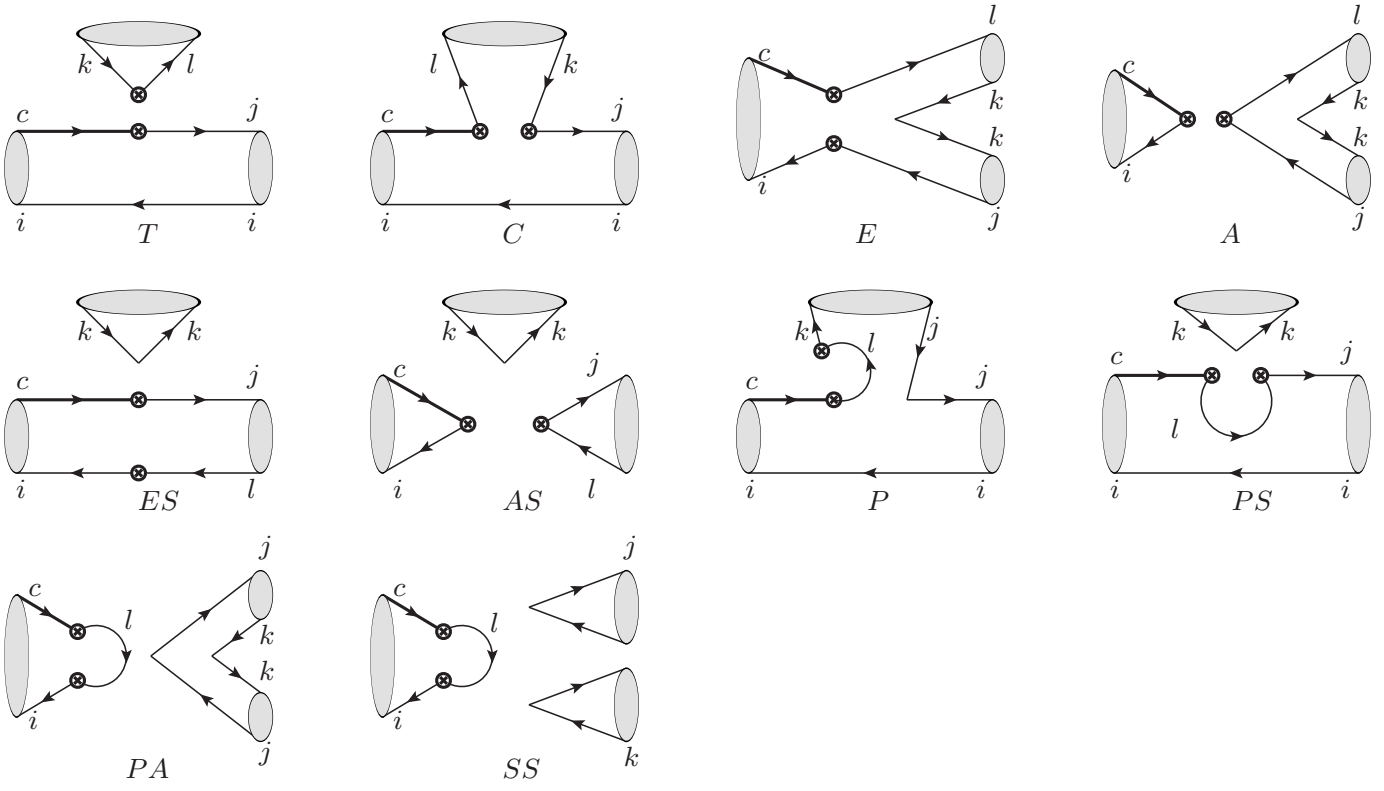}
  \caption{Topological diagrams contributing to the $D\to PP$ decay in the Standard Model.}\label{topo}
\end{figure}
In the $SU(3)$ picture, the pseudoscalar meson notet $|P^{i}_{j} \rangle$ is expressed as
\begin{eqnarray}\label{a1}
 |P^i_j\rangle =  \left( \begin{array}{ccc}
   \frac{1}{\sqrt 2} |\pi^0\rangle +  \frac{1}{\sqrt 6} |\eta_8\rangle,    & |\pi^+\rangle,  & |K^+\rangle \\
   | \pi^-\rangle, &   - \frac{1}{\sqrt 2} |\pi^0\rangle+ \frac{1}{\sqrt 6} |\eta_8\rangle,   & |K^0\rangle \\
   | K^- \rangle,& |\overline K^0\rangle, & -\sqrt{\frac{2}{3}}|\eta_8\rangle \\
  \end{array}\right) +  \frac{1}{\sqrt 3} \left( \begin{array}{ccc}
   |\eta_1\rangle,    & 0,  & 0 \\
    0, &  |\eta_1\rangle,   & 0 \\
   0, & 0, & |\eta_1\rangle \\
  \end{array}\right),
\end{eqnarray}
where $i$ is row index and $j$ is column index.
The vector meson nonet is
\begin{eqnarray}\label{v}
 |V\rangle ^i_j=  \left( \begin{array}{ccc}
   \frac{1}{\sqrt 2} |\rho^0\rangle+  \frac{1}{\sqrt 2} |\omega\rangle,    & |\rho^+\rangle,  & |K^{*+}\rangle \\
    |\rho^-\rangle, &   - \frac{1}{\sqrt 2} |\rho^0\rangle+ \frac{1}{\sqrt 2} |\omega\rangle,   & |K^{*0}\rangle \\
    |K^{*-}\rangle, & |\overline K^{*0}\rangle, & |\phi\rangle \\
  \end{array}\right).
\end{eqnarray}
The charmed meson anti-triplet state is
\begin{align}
|D^{i}\rangle = (|D^0\rangle, \,|D^+\rangle, \,|D_s^+\rangle).
\end{align}
The topological amplitude of the $D\to PP$ decay in the SM can be written as
\begin{align}\label{ha}
 \mathcal{A}(D \to P P)&=  T\,\cdot\, D_i  H^{lj}_kP^{i}_j  P^k_l + C\,\cdot\,D_i  H^{jl}_k  P^{i}_jP^k_l+E\,\cdot\,D_i  H^{il}_j P^j_k P^{k}_l+ A\,\cdot\,D_i H^{li}_j   P^j_k P^{k}_l\nonumber \\ &  +ES\,\cdot\,D_i  H^{ij}_{l}   P^{l}_j  P^k_k+AS\,\cdot\,D_i  H^{ji}_{l}  P^{l}_j  P^k_k +P\,\cdot\,D_i H^{kl}_{l}P^{i}_j   P^j_k + PS\,\cdot\,D_i  H^{jl}_{l} P^{i}_j  P^k_k   \nonumber\\& + PA\,\cdot\,D_i  H^{il}_{l}  P^j_k P^{k}_j + SS\,\cdot\,D_i   H^{il}_{l} P^{j}_j P^k_k.
\end{align}
If the index-contraction is understood as quark flowing,
each term in Eq.~\eqref{ha} is a topological diagram, see Fig.~\ref{topo}.
The first four diagrams, $T$, $C$, $E$ and $A$, have been analyzed in plenty of literature.
$ES$ and $AS$ are the singlet contributions which require multi-gluon exchanges.
The last four diagrams are quark-loop contributions.
More details about topological amplitudes can be found in Ref.~\cite{Wang:2020gmn}.

The amplitude $T$ is dominated by factorizable contribution, $T^{SD}$.
Here we use superscript "SD" to represent factorizable part compared with non-factorizable part.
In the amplitude $C$, the factorizable part $C^{SD}$ is also important since topology $C$ is the Fierz transformation of topology $T$.
In the naive factorization, the factorizable contributions in the $T$ and $C$ amplitudes are factorized into a product of decay constant $f_P$ and transition form factor $F_0^{D\rightarrow P}$:
\begin{align}
T^{SD} &=  \frac{G_F}{\sqrt{2}}V_{\rm CKM}a_{1}(\mu)f_{P_2}(m^2_D - m^2_{P_1})F_0^{D\rightarrow P_1}(m^2_{P_2}),\label{eq:TPP}
\\
C^{SD} &=  \frac{G_F}{\sqrt{2}}V_{\rm CKM}a_{2}(\mu)f_{P_2}(m^2_D - m^2_{P_1})F_0^{D\rightarrow P_1}(m^2_{P_2}),\label{eq:CPP}
\end{align}
with
\begin{align}
a_1(\mu) &= C_2(\mu) + \frac{C_1(\mu)}{N_c}, \qquad
a_2(\mu) = C_1(\mu) + \frac{C_2(\mu)}{N_c},
\end{align}
where $G_F$ denotes the Fermi coupling constant, $V_{\rm CKM}$ is the products of the Cabibbo-Kobayashi-Maskawa (CKM) matrix elements, $C_{1,2}$ are the Wilson coefficients.
In $T$ and $C$ diagrams, $P_1$ is the pseudoscalar meson transited from the $D$ decays and $P_2$ the emitted meson.
In the FSI framework, the non-factorable QCD effects can be modeled as an exchange of one particle between two particles generated from the short-distance tree emitted amplitudes, $T^{SD}$ and $C^{SD}$.
The $t$-channel FSI contribution forms a triangle diagram at hadron level, and can be derived from topological diagram via the topological-scattering diagram.
In the rest of this section, we try to establish the relation between the topological diagram, topological-scattering diagram and triangle diagram and analyze potential physical consequences.

\begin{figure}
  \centering
  \includegraphics[width=15cm]{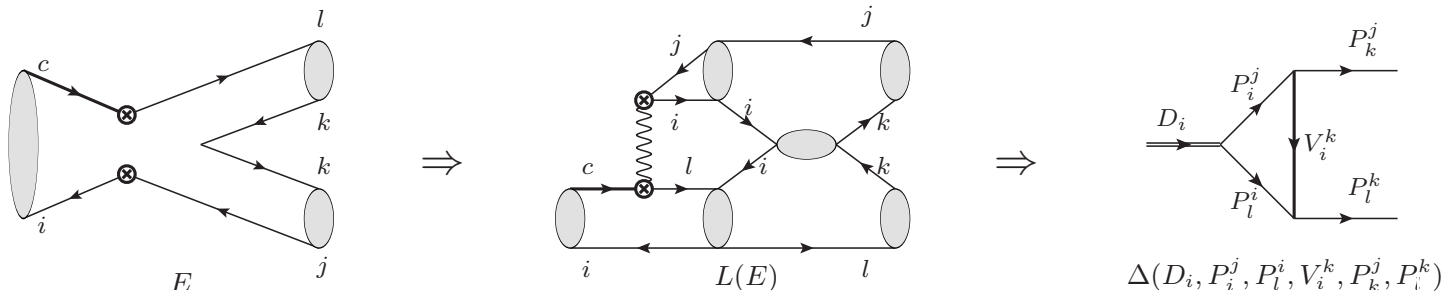}
  \caption{Topological diagram $\,\Rightarrow\,$ Topological-scattering diagram $\,\Rightarrow\,$ Triangle diagram in $T\Rightarrow E$ transition. The first right arrow represents the $t$-channel long-distance contributions induced by the short-distance $T$ amplitude in the $E$ diagram can be described by the topological-scattering diagram $L(E)$. The second right arrow represents the topological-scattering diagram $L(E)$ in quark level is equivalent to the triangle diagram $\Delta(D_i,P^j_i,P^i_l,V^k_i,P^j_k,P^k_l)$ at hadron level. }\label{E}
\end{figure}
\begin{figure}
  \centering
  \includegraphics[width=10cm]{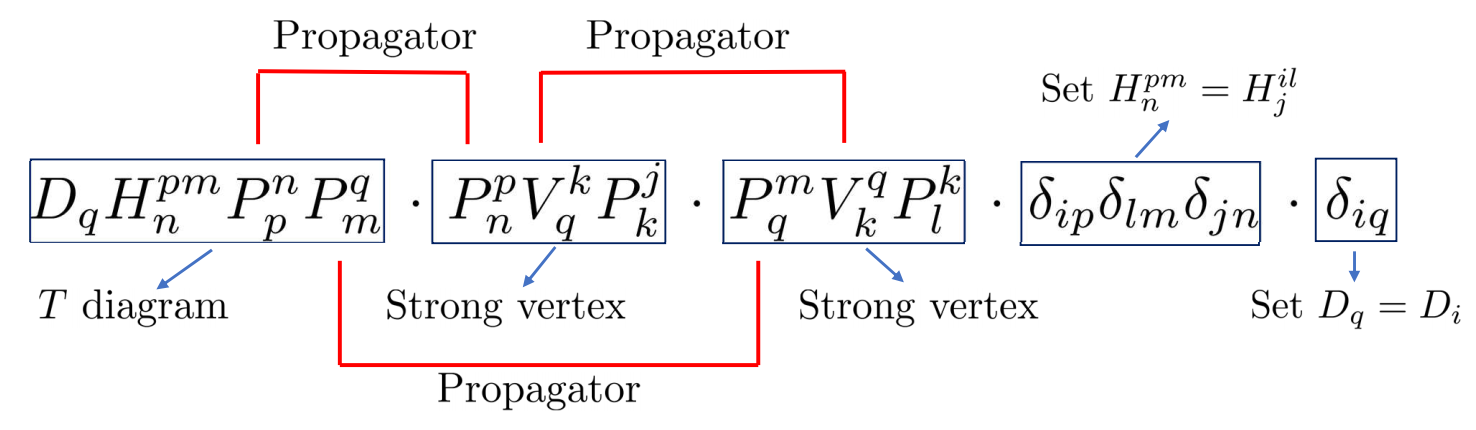}
  \caption{The triangle diagram constructed by the topological-scattering diagram in $T\,\Rightarrow\, E$ transition.}\label{relation}
\end{figure}

We express the topological-scattering diagram and triangle diagram in the tensor form, taking $E$ diagram as an example.
The rescattering contribution in $E$ diagram can be obtained by twisting quark lines from a short-distance $T$ diagram.
We call it "$T\Rightarrow E$" transition. The superscripts "SD" and "LD" are dropped for convenience.
The intermediate structure of $T\Rightarrow E$ transition is called topological-scattering diagram $L(E)$. And the topological-scattering diagram $L(E)$ forms a triangle diagram at hadron level, see Fig.~\ref{E}. In the tensor form of topological diagram, $T$ diagram is written as $D_q  H^{pm}_nP^{n}_p  P^q_m$, and $E$ diagram is written as $D_i  H^{il}_j P^j_k P^{k}_l$. Then the topological-scattering diagram of $T\Rightarrow E$ transition can be written as
\begin{align}\label{le}
L(E)[i,j,k,l]\,\,=\,\, D_qH^{pm}_nP^n_pP^q_m\,\cdot\,P^p_nV^k_qP^j_k\,\cdot\,P^m_qV^q_kP^k_l\,\cdot\,
\delta_{ip}\delta_{lm}\delta_{jn}\,\cdot\,\delta_{iq}.
\end{align}
$L(E)[i,j,k,l]$ can also be understood as a triangle diagram, see Fig.~\ref{relation}.
The $T$ diagram in the left is the weak vertex of $D_q\to P^n_pP^q_m$ in the triangle diagram. The two $PVP$ vertexes are two strong vertexes.
The index contractions of $P^n_p P^p_n$,  $P^q_m P^m_q$ and $V^k_q V^q_k$ are three meson propagators.
The kronecker symbols are used to set $H^{pm}_n=H^{il}_j$ and $D_q=D_i$.
Notice that we only consider the vector meson exchange here, i.e., $D\to PP \to PP$ via exchanging a vector meson.
For the processes such as $D\to VV \to PP$ via exchanging a pseudoscalar meson, the analysis is the similar to the case of $D\to PP \to PP$.
We shall not discuss it in this work.

\begin{figure}
  \centering
  \includegraphics[width=15cm]{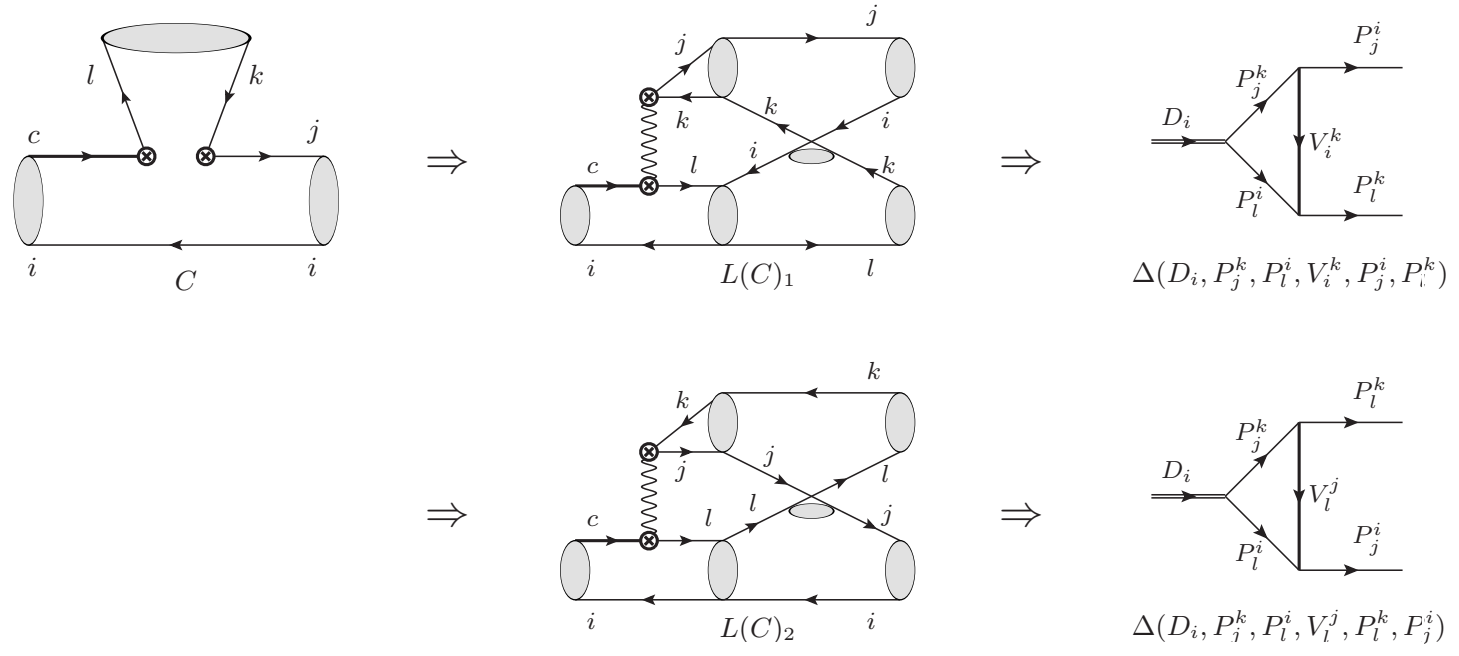}
  \caption{Topological diagram $\,\Rightarrow\,$ Topological-scattering diagram $\,\Rightarrow\,$ Triangle diagram in $T\,\Rightarrow\,C$ transition.}\label{C}
\end{figure}

Similar to the $E$ diagram, the topological-scattering diagram in $T\Rightarrow C$ triansition can be written as
\begin{align}
L(C)_1[i,j,k,l]\,\,&=\,\, D_qH^{pm}_nP^n_pP^q_m\,\cdot\,P^p_nV^n_iP^i_j\,\cdot\,P^m_qV^q_kP^k_l\,\cdot\,
\delta_{jp}\delta_{lm}\delta_{kn}\,\cdot\,\delta_{iq},\\
L(C)_2[i,j,k,l]\,\,&=\,\, D_qH^{pm}_nP^n_pP^q_m\,\cdot\,P^p_nV^l_pP^k_l\,\cdot\,P^m_qV^j_mP^i_j\,\cdot\,
\delta_{jp}\delta_{lm}\delta_{kn}\,\cdot\,\delta_{iq}.
\end{align}
The diagrams are presented in Fig.~\ref{C}.
Compared to the topological-scattering diagram $L(E)$, the two topological-scattering diagrams $L(C)_1$ and $L(C)_2$ have a cross in the above strong vertex.
This cross will contribute a minus sign in the triangle diagrams because of the commutator in chiral Lagrangian.
To make it clear, we analyze the strong vertexes of $\rho^0\to \pi^+\pi^-$ and $\omega\to \pi^+\pi^-$.
The chiral Lagrangian of $V\to PP$ decay is
\begin{align}
 \mathcal{L}_{VPP} = \frac{i}{\sqrt{2}}g_{VPP}Tr[V^{\mu}[P,\partial_\mu P]].
\end{align}
Inserting the pseudoscalar meson notet and vector meson nonet, one can get the strong vertexes of $\rho^0\to \pi^+\pi^-$ and $\omega\to \pi^+\pi^-$ are
\begin{align}\label{vpp}
 \mathcal{L}_{VPP} = & \frac{i}{\sqrt{2}}g_{VPP}[ ...+ (\frac{1}{\sqrt{2}}\rho^0+\frac{1}{\sqrt{2}}\omega)[\pi^+\partial_\mu\pi^-
 -\pi^-\partial_\mu\pi^+] +...\nonumber\\
 & + (-\frac{1}{\sqrt{2}}\rho^0+\frac{1}{\sqrt{2}}\omega)[\pi^-\partial_\mu\pi^+
 -\pi^+\partial_\mu\pi^-] +... ]\, ,
\end{align}
in which the first line is the contributions from $u\bar u$ constituent of $\rho^0/\omega$ and the second line is the contributions from $d\bar d$ constituent.
The $\omega\pi\pi$ vertexes from $u\bar u$ and $d\bar d$ cancel each other in the final chiral Lagrangian.
At the quark level, this cancelation can be described by Fig.~\ref{pipi}.
\begin{figure}
  \centering
  \includegraphics[width=12cm]{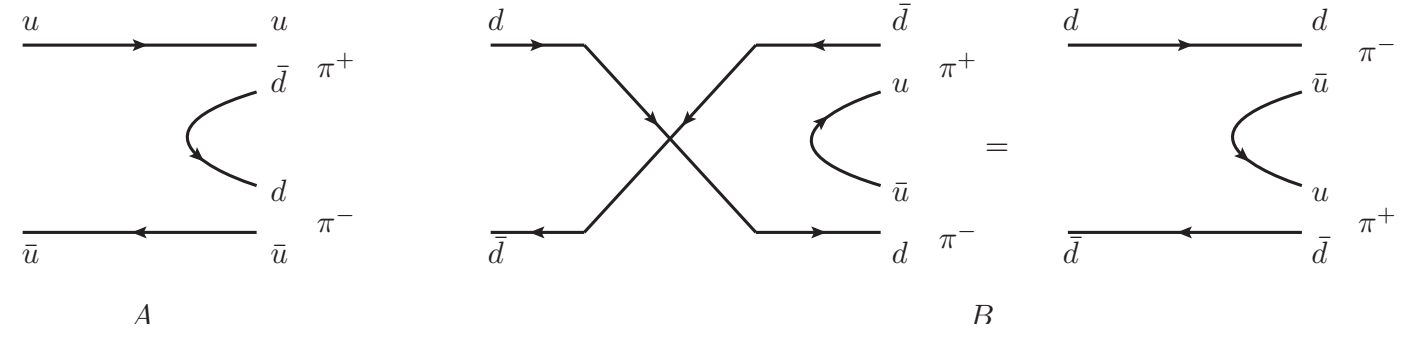}
  \caption{Strong vertexes of $\rho^0\to \pi^+\pi^-$ and $\omega\to \pi^+\pi^-$ .}\label{pipi}
\end{figure}
The first diagram in Fig.~\ref{pipi} is the $u\bar u$ contribution and the second diagram is the $d\bar d$ contribution.
The second diagram can be seen as an interchange of $\pi^+$ and $\pi^-$ in the first diagram.
Because of the commutator in chiral Lagrangian, the second diagram is opposite to the second diagram.
That is why $\omega\nrightarrow \pi^+\pi^-$.
For $\rho^0\to \pi^+\pi^-$ decay, there is a additional minus sign in the second diagram due to Eq.~\eqref{v} to get a non-zero amplitude.

\begin{figure}
  \centering
  \includegraphics[width=15cm]{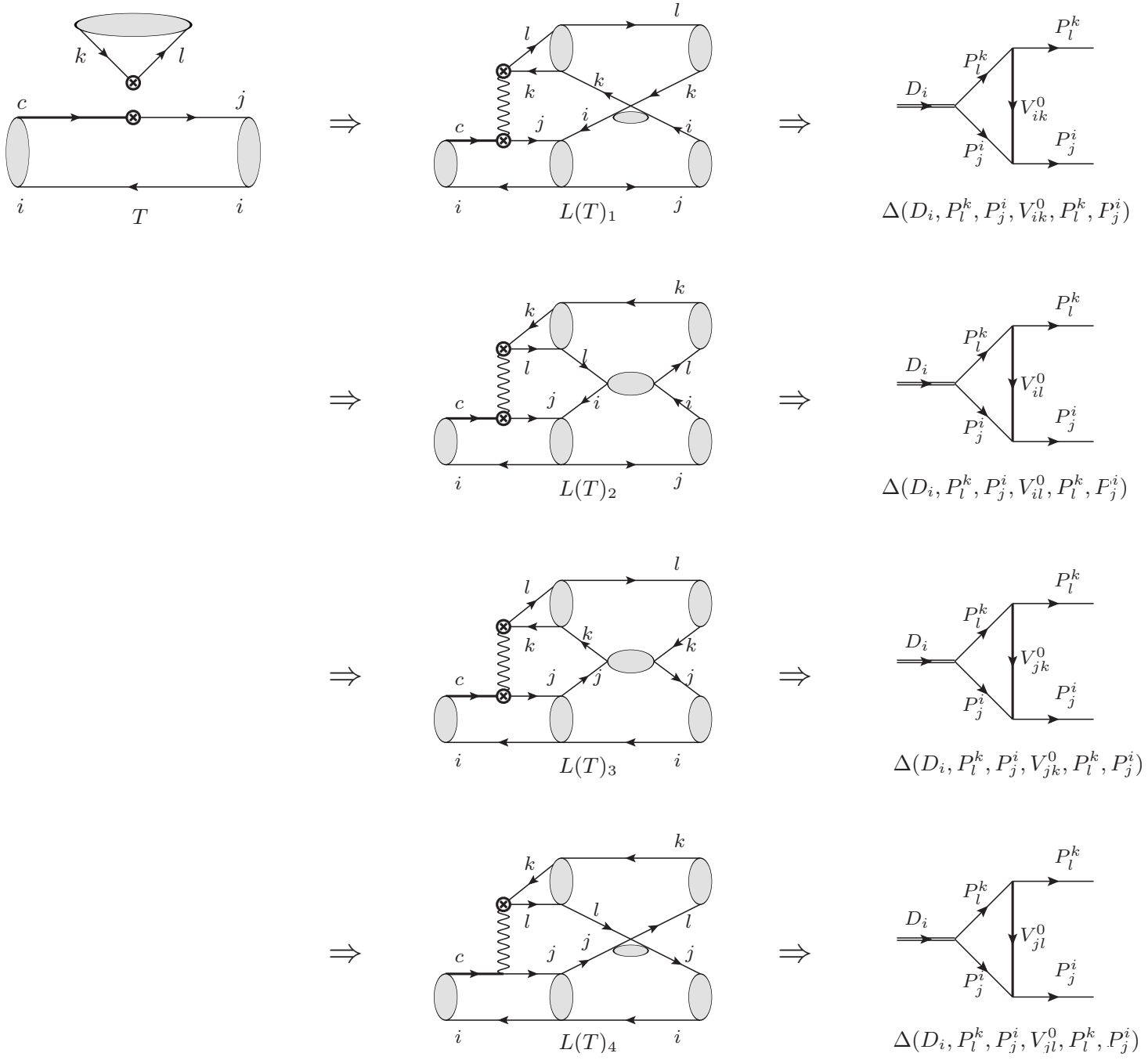}
  \caption{Topological diagram $\,\Rightarrow\,$ Topological-scattering diagram $\,\Rightarrow\,$ Triangle diagram in $T\,\Rightarrow\,T$ transition.}\label{T}
\end{figure}

\begin{figure}
  \centering
  \includegraphics[width=10cm]{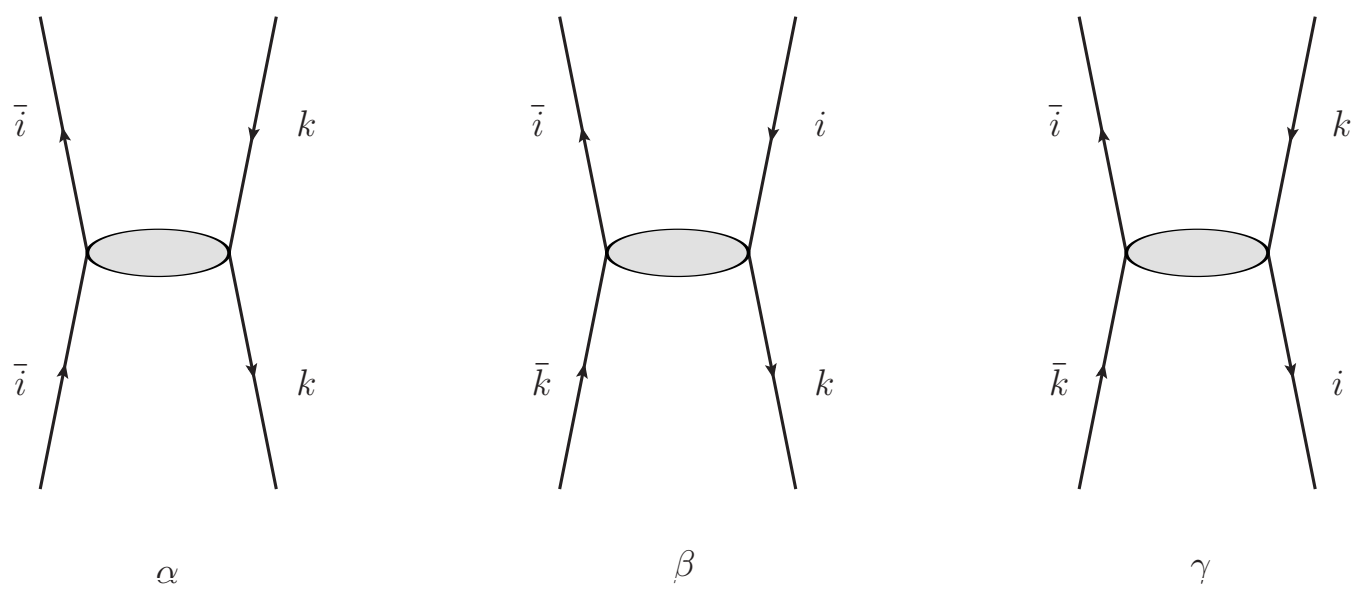}
  \caption{Comparison between the meson propagator constructed by an invariant tensor ($\alpha$) and two dyadic tensors ($\beta$ and $\gamma$).}\label{pauli}
\end{figure}

The long-distance contributions in $T$ diagram can also be modeled as meson-meson scattering between two mesons emitted from short-distance $T$ diagram.
Similar to the $E$ and $C$ diagrams, the topological-scattering diagram in $T\Rightarrow T$ transition can be written as
\begin{align}\label{T1}
L(T)_1[i,j,k,l]\,\,&=\,\, D_qH^{pm}_nP^n_pP^q_m\,\cdot\,P^p_nV^n_kP^k_l\,\cdot\,P^m_qV^q_iP^i_j\,\cdot\,
\delta_{lp}\delta_{jm}\delta_{kn}\,\cdot\,\delta_{iq},\\
L(T)_2[i,j,k,l]\,\,&=\,\, D_qH^{pm}_nP^n_pP^q_m\,\cdot\,P^p_nV^l_pP^k_l\,\cdot\,P^m_qV^q_iP^i_j\,\cdot\,
\delta_{lp}\delta_{jm}\delta_{kn}\,\cdot\,\delta_{iq},\\
L(T)_3[i,j,k,l]\,\,&=\,\, D_qH^{pm}_nP^n_pP^q_m\,\cdot\,P^p_nV^n_kP^k_l\,\cdot\,P^m_qV^j_mP^i_j\,\cdot\,
\delta_{lp}\delta_{jm}\delta_{kn}\,\cdot\,\delta_{iq},\\ \label{T4}
L(T)_4[i,j,k,l]\,\,&=\,\, D_qH^{pm}_nP^n_pP^q_m\,\cdot\,P^p_nV^l_pP^k_l\,\cdot\,P^m_qV^j_mP^i_j\,\cdot\,
\delta_{lp}\delta_{jm}\delta_{kn}\,\cdot\,\delta_{iq}.
\end{align}
The diagrams are presented in Fig.~\ref{T}.
The vector propagators in the Eqs.~\eqref{T1}$\sim$ \eqref{T4} are constructed by two dyadic tensors \cite{Morse1} rather than an invariant tensor such as $V^k_i V^i_k$ in $T\Rightarrow E$ transition.
For example, the vector propagator in $L(T)_1$ is expressed as $V^n_kV^q_i$. Considering $i=q$ and $k=n$, it becomes $V^k_kV^i_i$.
The upper and lower indies in $V^k_k$ and $V^i_i$ do not contract each other like a singlet.
$V^k_kV^i_i$ can be seen as a neutral vector propagator, such as $\rho^0$, $\omega$ ... etc.
In order to clarify the difference between dyadic tensor and invariant tensor, we show the invariant tensor ($\alpha$) and dyadic tensor ($\beta$ and $\gamma$) in Fig.~\ref{pauli}.
In $\alpha$ diagram, there are two quark lines, $\bar i$ and $k$.
In $\beta$ diagram, there are four quark lines, $i$, $\bar i$, $k$, $\bar k$.
If $i=k$, we can interchange the position of $i$ and $k$ to transform $\beta$ diagram to $\gamma$ diagram. According to the Pauli exclusion principle, interchanging two fermions produces a minus sign, thus $\beta+\gamma=0$. So $\beta$ diagram is a neutral vector propagator only in the case of $i\neq k$.

For the completeness of our theoretical framework, we list all the tensor structures in $T\Rightarrow A,\,ES,\,AS,\,P,\,PS,\,PA,\,SS$ transitions. \\
$T\Rightarrow A$:
\begin{align}
L(A)_1[i,j,k,l]\,\,&=\,\, D_qH^{pm}_nP^n_pP^q_m\,\cdot\,P^p_nV^k_pP^j_k\,\cdot\,P^r_kV^l_rP^k_l\,\cdot\,
\delta_{lp}\delta_{im}\delta_{jn}\,\cdot\,\delta_{iq}\,\cdot\,\delta_{kr},\\
L(A)_2[i,j,k,l]\,\,&=\,\, D_qH^{pm}_nP^n_pP^q_m\,\cdot\,P^p_nV^n_kP^k_l\,\cdot\,P^k_rV^r_jP^j_k\,\cdot\,
\delta_{lp}\delta_{im}\delta_{jn}\,\cdot\,\delta_{iq}\,\cdot\,\delta_{kr},\\
L(A)_3[i,j,k,l]\,\,&=\,\, D_qH^{pm}_nP^n_pP^q_m\,\cdot\,P^p_nV^k_pP^j_k\,\cdot\,P^l_rV^r_kP^k_l\,\cdot\,
\delta_{lp}\delta_{im}\delta_{jn}\,\cdot\,\delta_{iq}\,\cdot\,\delta_{lr},\\
L(A)_4[i,j,k,l]\,\,&=\,\, D_qH^{pm}_nP^n_pP^q_m\,\cdot\,P^p_nV^n_kP^k_l\,\cdot\,P^r_jV^k_rP^j_k\,\cdot\,
\delta_{lp}\delta_{im}\delta_{jn}\,\cdot\,\delta_{iq}\,\cdot\,\delta_{jr},\\
L(A)_5[i,j,k,l]\,\,&=\,\, D_qH^{pm}_nP^n_pP^q_m\,\cdot\,P^p_nV^k_pP^i_j\,\cdot\,P^m_qV^l_kP^k_l\,\cdot\,
\delta_{lp}\delta_{im}\delta_{jn}\,\cdot\,\delta_{iq},\\
L(A)_6[i,j,k,l]\,\,&=\,\, D_qH^{pm}_nP^n_pP^q_m\,\cdot\,P^p_nV^n_kP^k_l\,\cdot\,P^m_qV^k_jP^j_k\,\cdot\,
\delta_{lp}\delta_{im}\delta_{jn}\,\cdot\,\delta_{iq}.
\end{align}
$T\Rightarrow ES$:
\begin{align}
L(ES)_1[i,j,k,l]\,\,&=\,\, D_qH^{pm}_nP^n_pP^q_m\,\cdot\,P^p_nV^j_pP^l_j\,\cdot\,P^m_qV^q_mP^k_k\,\cdot\,
\delta_{ip}\delta_{jm}\delta_{ln}\,\cdot\,\delta_{iq},\\
L(ES)_2[i,j,k,l]\,\,&=\,\, D_qH^{pm}_nP^n_pP^q_m\,\cdot\,P^p_nV^n_pP^k_k\,\cdot\,P^m_qV^q_lP^l_j\,\cdot\,
\delta_{ip}\delta_{jm}\delta_{ln}\,\cdot\,\delta_{iq}.
\end{align}
$T\Rightarrow AS$:
\begin{align}
L(AS)_1[i,j,k,l]\,\,&=\,\, D_qH^{pm}_nP^n_pP^q_m\,\cdot\,P^p_nV^n_lP^l_j\,\cdot\,P^s_rV^k_sP^r_k\,\cdot\,
\delta_{jp}\delta_{im}\delta_{ln}\,\cdot\,\delta_{iq}\,\cdot\,\delta_{kr}
\,\cdot\,\delta_{ks},\\
L(AS)_2[i,j,k,l]\,\,&=\,\, D_qH^{pm}_nP^n_pP^q_m\,\cdot\,P^p_nV^j_pP^l_j\,\cdot\,P^s_rV^k_sP^r_k\,\cdot\,
\delta_{jp}\delta_{im}\delta_{ln}\,\cdot\,\delta_{iq}\,\cdot\,\delta_{kr}
\,\cdot\,\delta_{ks},\\
L(AS)_3[i,j,k,l]\,\,&=\,\, D_qH^{pm}_nP^n_pP^q_m\,\cdot\,P^p_nV^n_lP^l_j\,\cdot\,P^k_sV^s_rP^r_k\,\cdot\,
\delta_{jp}\delta_{im}\delta_{ln}\,\cdot\,\delta_{iq}\,\cdot\,\delta_{kr}
\,\cdot\,\delta_{ks},\\
L(AS)_4[i,j,k,l]\,\,&=\,\, D_qH^{pm}_nP^n_pP^q_m\,\cdot\,P^p_nV^j_pP^l_j\,\cdot\,P^k_sV^s_rP^r_k\,\cdot\,
\delta_{jp}\delta_{im}\delta_{ln}\,\cdot\,\delta_{iq}\,\cdot\,\delta_{kr}
\,\cdot\,\delta_{ks},\\
L(AS)_5[i,j,k,l]\,\,&=\,\, D_qH^{pm}_nP^n_pP^q_m\,\cdot\,P^p_nV^n_lP^l_j\,\cdot\,P^m_qV^k_rP^r_k\,\cdot\,
\delta_{jp}\delta_{im}\delta_{ln}\,\cdot\,\delta_{iq}\,\cdot\,\delta_{kr},\\
L(AS)_6[i,j,k,l]\,\,&=\,\, D_qH^{pm}_nP^n_pP^q_m\,\cdot\,P^p_nV^j_pP^l_j\,\cdot\,P^m_qV^k_rP^r_k\,\cdot\,
\delta_{jp}\delta_{im}\delta_{ln}\,\cdot\,\delta_{iq}\,\cdot\,\delta_{kr},\\
L(AS)_7[i,j,k,l]\,\,&=\,\, D_qH^{pm}_nP^n_pP^q_m\,\cdot\,P^p_nV^n_lP^l_j\,\cdot\,P^m_qV^q_mP^k_k\,\cdot\,
\delta_{jp}\delta_{im}\delta_{ln}\,\cdot\,\delta_{iq},\\
L(AS)_8[i,j,k,l]\,\,&=\,\, D_qH^{pm}_nP^n_pP^q_m\,\cdot\,P^p_nV^j_pP^l_j\,\cdot\,P^m_qV^q_mP^k_k\,\cdot\,
\delta_{jp}\delta_{im}\delta_{ln}\,\cdot\,\delta_{iq},\\
L(AS)_9[i,j,k,l]\,\,&=\,\, D_qH^{pm}_nP^n_pP^q_m\,\cdot\,P^p_nV^n_pP^k_k\,\cdot\,P^m_qV^j_lP^l_j\,\cdot\,
\delta_{jp}\delta_{im}\delta_{ln}\,\cdot\,\delta_{iq}.
\end{align}
$T\Rightarrow P$:
\begin{align}
L(P)[i,j,k,l]\,\,&=\,\, D_qH^{pm}_nP^n_pP^q_m\,\cdot\,P^p_nV^n_jP^j_k\,\cdot\,P^m_qV^j_mP^i_j\,\cdot\,
\delta_{kp}\delta_{lm}\delta_{ln}\,\cdot\,\delta_{iq}.
\end{align}
$T\Rightarrow PS$
\begin{align}
L(PS)_1[i,j,k,l]\,\,&=\,\, D_qH^{pm}_nP^n_pP^q_m\,\cdot\,P^p_nV^n_iP^i_j\,\cdot\,P^m_qV^q_mP^k_k\,\cdot\,
\delta_{jp}\delta_{lm}\delta_{ln}\,\cdot\,\delta_{iq},\\
L(PS)_2[i,j,k,l]\,\,&=\,\, D_qH^{pm}_nP^n_pP^q_m\,\cdot\,P^p_nV^n_pP^k_k\,\cdot\,P^m_qV^j_mP^i_j\,\cdot\,
\delta_{jp}\delta_{lm}\delta_{ln}\,\cdot\,\delta_{iq}.
\end{align}
$T \Rightarrow PA$: no contraction.\\
$T\Rightarrow SS$:
\begin{align}
L(SS)[i,j,k,l]\,\,&=\,\, D_qH^{pm}_nP^n_pP^q_m\,\cdot\,P^p_nV^n_pP^j_j\,\cdot\,P^m_qV^q_mP^k_k\,\cdot\,
\delta_{ip}\delta_{lm}\delta_{ln}\,\cdot\,\delta_{iq}.
\end{align}
Notice that not all the above tensor structures can be understood as triangle diagrams.
For example, $L(ES)_1$ cannot form a triangle diagram at hadron level because $P^m_qV^q_mP^k_k$ is not an effective strong vertex like Fig.~\ref{pipi}.
\begin{figure}
  \centering
  \includegraphics[width=15cm]{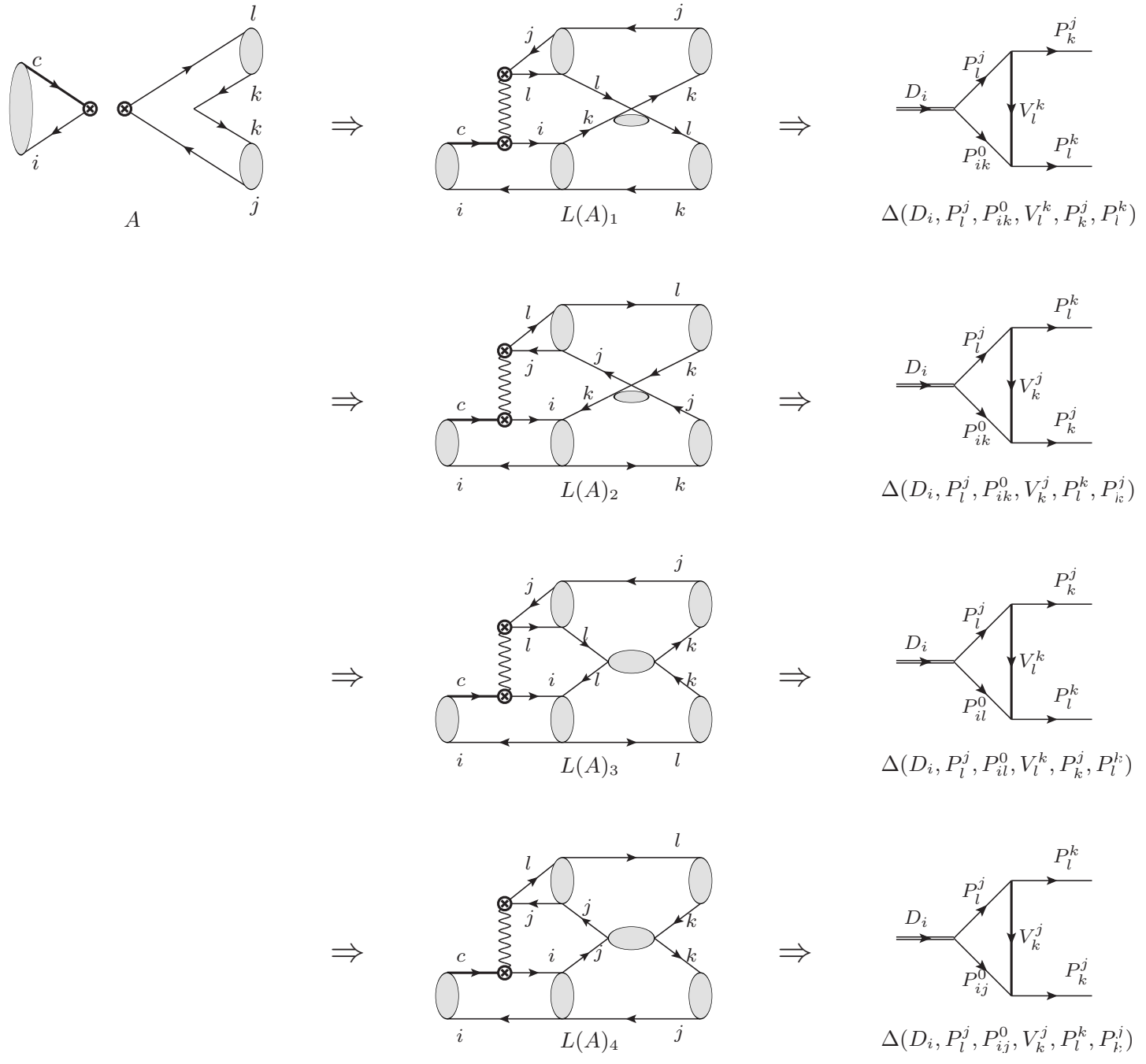}
  \caption{Topological diagram $\,\Rightarrow\,$ Topological-scattering diagram $\,\Rightarrow\,$ Triangle diagram in $T\,\Rightarrow\,A$ transition.}\label{A}
\end{figure}
\begin{figure}
  \centering
  \includegraphics[width=15cm]{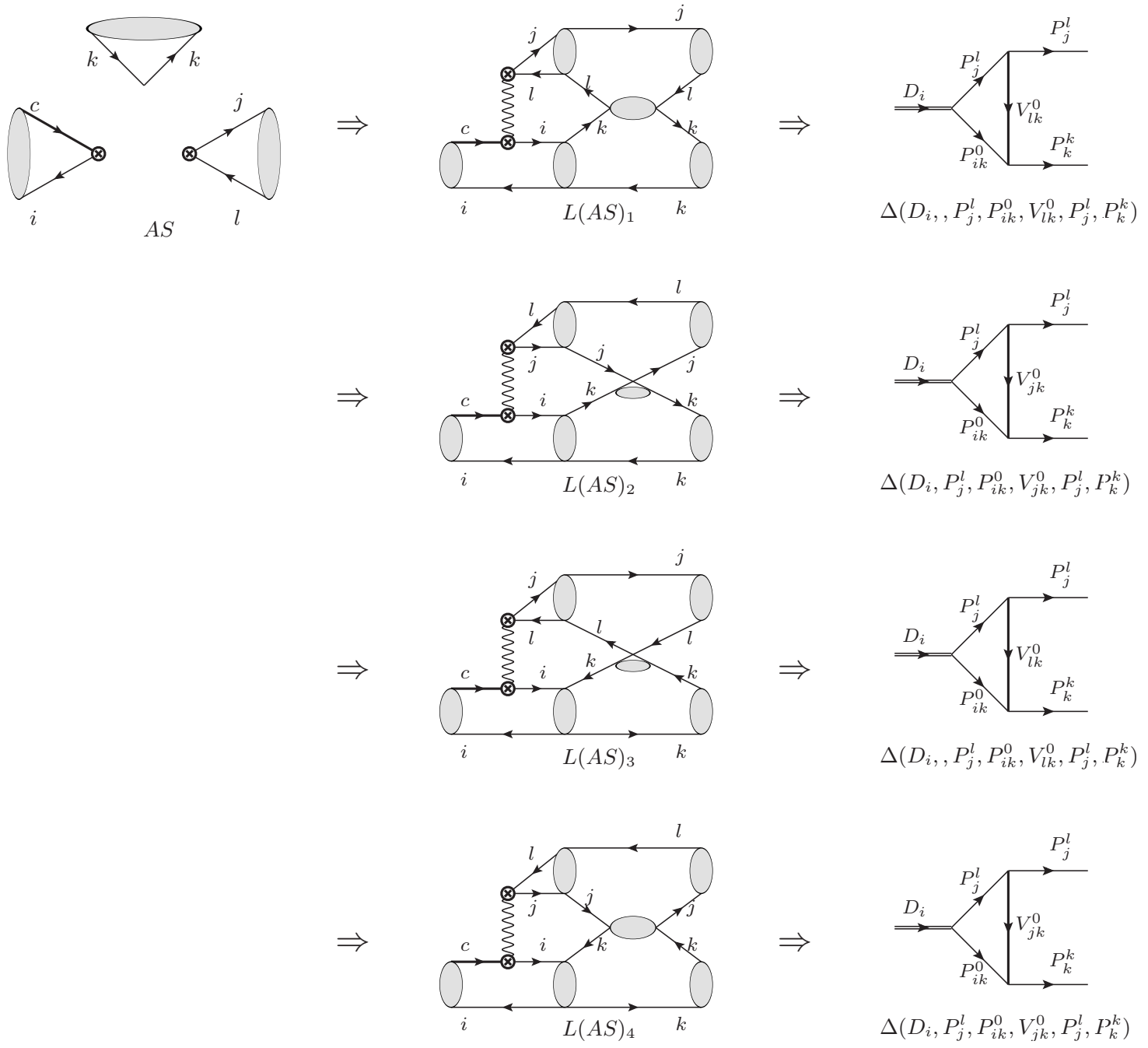}
  \caption{Topological diagram $\,\Rightarrow\,$ Topological-scattering diagram $\,\Rightarrow\,$ Triangle diagram in $T\,\Rightarrow\,AS$ transition.}\label{AS}
\end{figure}
\begin{figure}
  \centering
  \includegraphics[width=15cm]{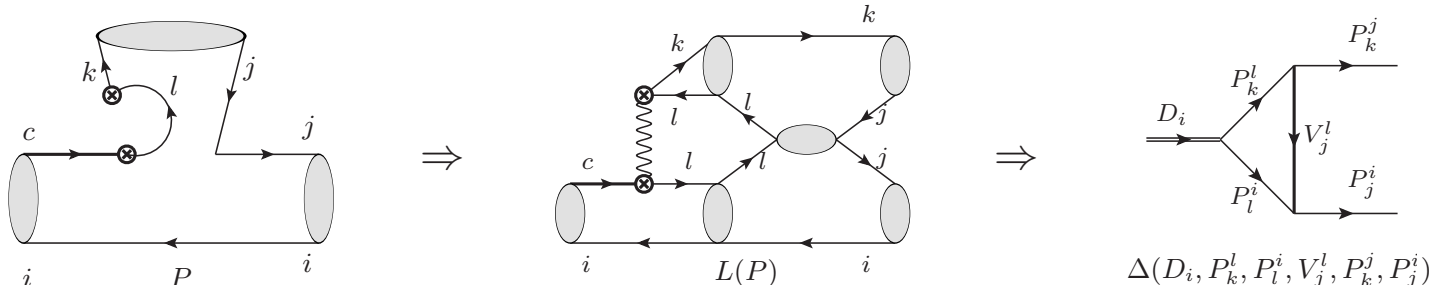}
  \caption{Topological diagram $\,\Rightarrow\,$ Topological-scattering diagram $\,\Rightarrow\,$ Triangle diagram in $T\,\Rightarrow\,P$ transition.}\label{LP}
\end{figure}
The triangle diagrams constructed by the topological-scattering diagrams in $T\Rightarrow A$, $T\Rightarrow AS$ and $T\Rightarrow P$ transitions are presented in Figs.~\ref{A}, \ref{AS} and \ref{LP} respectively.

The triangle diagrams in the $T\Rightarrow AS$ transition cancel each other.
In Fig.~\ref{AS}, the triangle diagrams derived by topological-scattering diagrams $L(AS)_1$ and $L(AS)_3$ are the same.
But there is a cross in the topological-scattering diagram $L(AS)_3$, which leads to a minus in the triangle diagram, and hence $L(AS)_1+L(AS)_3=0$.
The same situation also appears in the topological-scattering diagrams $L(AS)_2$ and $L(AS)_4$. Thereby, the triangle diagrams associated with $T\Rightarrow AS$ transition do not contribute.

\begin{figure}
  \centering
  \includegraphics[width=12cm]{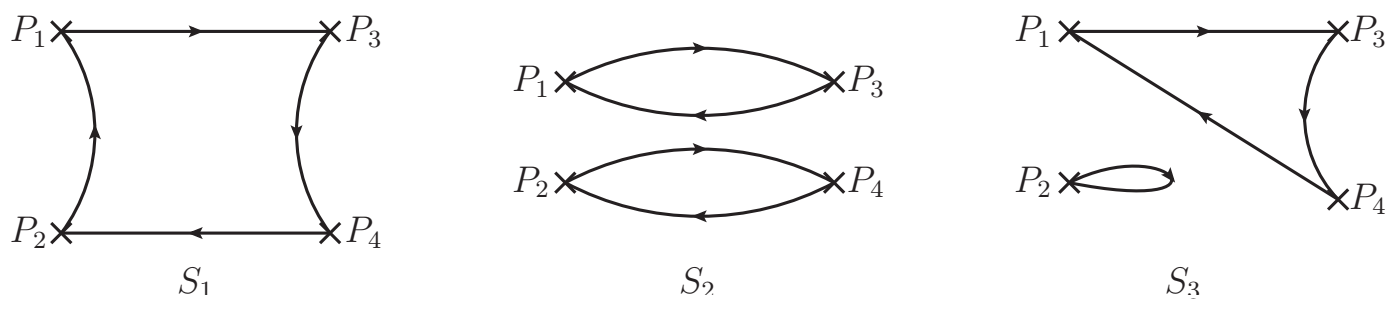}
  \caption{Sketches of meson-meson scattering in the rescattering contributions of $C$, $E$, $P$ diagrams (left), $T$ diagram (middle) and $A$ diagram (right).}\label{scatter}
\end{figure}
The meson-meson scattering in the $C$, $E$ and $P$ diagrams is sketched by $S_1$ in Fig.~\ref{scatter}. And the scattering in the $T$ diagram and $A$ diagram are sketched by $S_2$ and $S_3$, respectively.
Specifically, diagram $S_1$ in Fig.~\ref{scatter} is the meson-meson scattering in $T \Rightarrow P$ transition, Fig.~\ref{LP}.
Flipping the direction of quark lines of $S_1$, one can get the meson-meson scattering in $T \Rightarrow E$ transition, Fig.~\ref{E}.
Twisting the quark lines of $S_1$ and its flipping to interchange the position of $P_3$ and $P_4$, one can get the meson-meson scattering in $T \Rightarrow C$ transition, Fig.~\ref{C}.
Thereby, all the four possible topological structures of $S_1$ are  included in the topological-scattering diagrams in $T \Rightarrow E,\,C,\,P$ transitions. For the $S_2$ diagram in Fig.~\ref{scatter}, we can tab the four quark lines with $1$, $2$, $3$, $4$ orderly.
The intermediate vector propagators of the four topological-scattering diagrams in Fig.~\ref{T} are formed by the $24$, $14$, $23$, $13$ quark lines of $S_2$ in turn.
All the four possible topological structures of $S_2$ are included in the four topological-scattering diagrams in $T \Rightarrow T$ transition. The $S_3$ diagram in Fig.~\ref{scatter} is the meson-meson scattering of the fourth topological-scattering diagram in Fig.~\ref{A}.
Flipping the direction of quark lines of the above sub-diagram of $S_3$, we get the meson-meson scattering of the third topological-scattering diagram in Fig.~\ref{A}.
Twisting the quark lines of the above sub-diagram of $S_3$ and its flipping to interchange the position of $P_3$ and $P_4$, we get the meson-meson scattering of the first two topological-scattering diagram in Fig.~\ref{A}.
And again, all the four possible structures of $S_3$ are included in the four topological-scattering diagrams in $T \Rightarrow A$ transition.
In the end, the topological-scattering diagrams in $T \Rightarrow T$, $T \Rightarrow C$, $T \Rightarrow E$, $T \Rightarrow A$ and $T \Rightarrow P$ transitions cover all the twelve possible topological structures of meson-meson scattering without repetition.
So Fig.~\ref{scatter} helps us make clear the completeness of topological-scattering diagram.

\begin{figure}
  \centering
  \includegraphics[width=15cm]{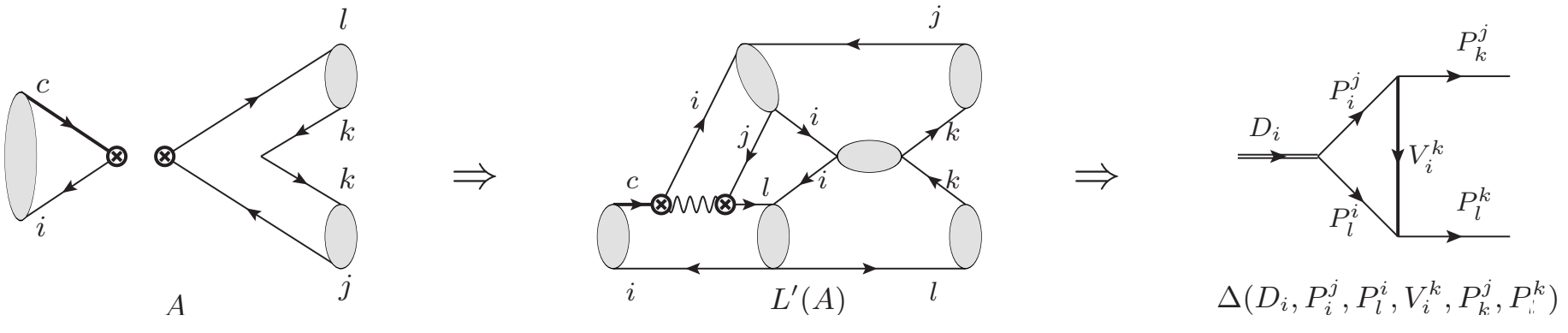}
  \caption{Topological diagram $\,\Rightarrow\,$ Topological-scattering diagram $\,\Rightarrow\,$ Triangle diagram in $C\,\Rightarrow\,A$ transition.}\label{Ac}
\end{figure}

\begin{figure}
  \centering
  \includegraphics[width=3.7cm]{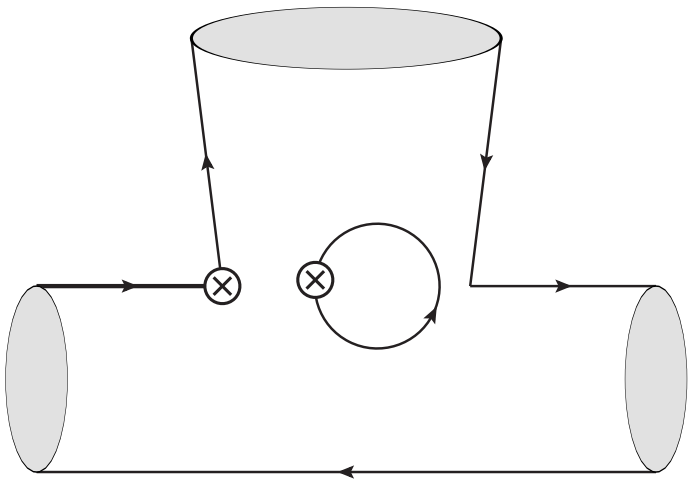}
  \caption{$P^\prime$ diagram, the Fierz transformation of $P$ diagram.}\label{px}
\end{figure}

The rescattering contributions can also be constructed by the two mesons emitted from the short-distance $C$ diagram, $C^{SD}$. In the topological-scattering diagram, it is equivalent to use the short-distance $C$ diagram to replace the short-distance $T$ diagram.
For example, the topological-scattering diagram of $C \Rightarrow A$ transition $L^\prime(A)$ can be obtained from the topological-scattering diagram of $T \Rightarrow E$ transition $L(E)$, see Fig.~\ref{Ac}.
The relations between the topological-scattering diagrams arisen from $C^{SD}$ and $T^{SD}$ are summarized to be
\begin{align}\label{csd}
& L^{\prime}(C)_i = \frac{C^{SD}}{T^{SD}}L(T)_i, \qquad L^{\prime}(A) = \frac{C^{SD}}{T^{SD}}L(E),\qquad L^{\prime}(T)_i = \frac{C^{SD}}{T^{SD}}L(C)_i, \nonumber\\
& L^{\prime}(E)_i = \frac{C^{SD}}{T^{SD}}L(A)_i, \qquad L^{\prime}(P^\prime) = \frac{C^{SD}}{T^{SD}}L(P),
\end{align}
in which $P^\prime$ diagram is the Fierz transformation of $P$ diagram, see Fig.~\ref{px}.
In the SM, $P^\prime$ diagram is zero if the tree operators $O_1$ or $Q_2$ is inserted into its weak vertex. The ratio $C^{SD}/T^{SD}$ in Eq.~\eqref{csd} is suppressed by the color factor at charm scale \cite{Li:2012cfa}
\begin{align}
\frac{a_2(m_c)}{a_1(m_c)}=\frac{C_1(m_c)+C_2(m_c)/3}{C_2(m_c)+C_1(m_c)/3}\sim \mathcal O(10^{-2\sim -1}).
\end{align}
Thus the rescattering contributions arisen from $C^{SD}$ are about one order smaller than ones arisen from $T^{SD}$. The factorizable contributions of $E$ and $A$ amplitudes are down by the helicity suppression, and the short-distance contribution of $P$ amplitude is roughly estimated to be $P^{SD}/T^{SD}\sim\alpha_s(m_c)/\pi\sim 0.1$. The rescattering contributions arisen from $E^{SD}$, $A^{SD}$ and $P^{SD}$ are negligible compared to the ones arisen from $T^{SD}$.

The $ES$, $AS$, $PS$, $PA$ and $SS$ diagrams can be divided into several unconnected parts by cutting off gluon propagators.
They are suppressed by the OZI rule \cite{OZI1,OZI2,OZI3} and can not be constructed by scattering of the two mesons generated by a short-distance diagram.
On the contrary, the $T$, $C$, $E$, $A$ and $P$ diagrams cannot be divided into unconnected sub-diagram by cutting off gluon propagators and hence their long-distance contributions could be modeled into meson-meson scattering.

For convenience to the subsequent analysis, we factor out the coefficient of constituent quarks of meson in the triangle diagram.
For example, the $u\bar u$ constituent for pseudoscalar meson can be written as $1/\sqrt{2}\pi^0+1/\sqrt{6}\eta_8+1/\sqrt{3}\eta_1$.
If a $\pi^0$ with $u\bar u$ constituent emits as a final state, the factor $1/\sqrt{2}$ is multiplied before the triangle diagram.
If the $u\bar u$ constituent acts as a propagator, the propagator should be written as $\pi^0/2+\eta_8/6+\eta_1/3$ since there are two effective vertexes at triangle diagram.
In this way, the propagators of charmed meson anti-triplet state, pseudoscalar meson notet and vector meson nonet and the coupling vertex between them are separately equivalent in the triangle diagram under the $SU(3)_F$ symmetry.

Under the $SU(3)_F$ symmetry, the rescattering contributions arisen from $T^{SD}$ in the $T$ and $A$ diagrams are zero because of the neutral propagators constructed by dyadic tensors.
For example, if $i=u$, $k=d$ in $L(T)_1$, the vector propagator $V^0_{ik}$ is $-\rho^0/2+\omega/2$.
Under the $SU(3)_F$ symmetry, $\rho^0 = \omega$, and hence $L(T)_1=0$.
This situation also appears in other topological-scattering diagrams of $T$ and $A$ diagrams.
Thereby, the rescattering contributions arisen from $T^{SD}$ in the $T$ and $A$ diagrams are only induced by the $SU(3)_F$ breaking effects.
The dyadic tensors do not appear in $T\Rightarrow E$, $T\Rightarrow C$ and $T\Rightarrow P$ transitions.
Considering the $C^{SD}$ amplitude is one order smaller than the $T^{SD}$ amplitude, the rescattering contributions in the $T$ and $A$ diagrams are smaller than $C$, $E$ and $P$ diagrams.
The different size between rescattering contributions in the $T$, $A$ diagrams and the $C$, $E$ and $P$ diagrams can be understood in the large $N_c$ expansion. In Fig.~\ref{scatter}, $S_1$ is a color singlet. It is at order of $(1/\sqrt{N_c})^4N_c=1/N_c$ in the large $N_c$ expansion.
$S_2$ and $S_3$ include two color singlets. They are at order of $(1/\sqrt{N_c})^4(N_c)^2/(N_c)^2=1/N_c^2$ in the large $N_c$ expansion.

All triangle diagrams are equivalent under the $SU(3)_F$ symmetry.
considering that there are two topological-scattering diagrams in $T\Rightarrow C$ transition and one topological-scattering diagram in $T\Rightarrow E$ and $T\Rightarrow P$ transitions as well as the minus sign induced by the cross of quark lines in the topological-scattering diagrams in $T\Rightarrow C$ transition, we conclude that
\begin{align}\label{re}
 L(C):L(E):L(P)=-2:1:1.
\end{align}
As we mentioned above, the short-distance amplitudes $C^{SD}$, $E^{SD}$, $A^{SD}$ and $P^{SD}$ are estimated to be one order smaller than $T^{SD}$.
Eq.~\eqref{re} will not be obviously broken if other rescattering contributions are included.
If the $s$-channel one particle exchange, i.e., the resonance contribution, is considered, Eq.~\eqref{re} is broken.
But the proportional relation is still significant since it gives a rough estimation of the long-distance contributions.

In Refs.~\cite{Li:2002pj,Ablikim:2002ep}, the rescattering contributions of $C$ and $E$ diagrams are estimated to be the same order with the factorizable amplitude $T^{SD}$.
From Eq.~\eqref{re}, one can find the rescattering contribution in $P$ diagram is at the same order with $C$ and $E$ diagrams.
So the penguin amplitudes are the same order with the tree amplitudes.
It is consistent with the large penguin diagram extracted from CP violation in charm decay \cite{Aaij:2019kcg,Grossman:2019xcj}.
In Ref.~\cite{Cheng:2012xb}, the authors assumed $PE=E$ and predicted that $\Delta A_{CP}=(-1.39\pm 0.04)\times 10^{-3}$ or $(-1.51\pm 0.04)\times 10^{-3}$, very close to the experimental value, $\Delta A_{CP} = (-1.54\pm 0.29)\times 10^{-3}$ \cite{Aaij:2019kcg}.
In fact, the assumptions that $PE=E$, $P=E$ or $P+PE=E$ will give similar results because $P$ and $PE$ defined in \cite{Cheng:2012xb} always appear as $P+PE$.
In Ref.~\cite{Wang:2020gmn}, we conclude that topology $P$ defined in this work is identical to $P+PE$ defined in \cite{Cheng:2012xb}. So $L(P)=L(E)$ is consistent with the hypothesis proposed in \cite{Cheng:2012xb}.

\section{Application in $D$ meson decay}\label{cdecay}
In this section, we take the $D\to K\pi$ and $D\to \pi\pi$ decays as examples to illustrate the reliability of the theoretical framework.
We only analyze the rescattering contributions arisen from $T^{SD}$ since the rescattering contributions arisen from other short-distance amplitudes are suppressed compared to the ones arisen from $T^{SD}$.

\subsection{$D\to K\pi$ decay}\label{dkpi}
The decay modes $D^+\to \overline K^0\pi^+$, $D^0\to K^-\pi^+$ and $D^0\to \overline K^0\pi^0$ have following Isospin relation
\begin{align}\label{kpi}
 \mathcal{A}(D^+\to \overline K^0\pi^+)-\mathcal{A}(D^0\to K^-\pi^+)=\sqrt{2}\mathcal{A}(D^0\to \overline K^0\pi^0).
\end{align}
The topological amplitude of $D^0\to K^-\pi^+$ decay is $T+E$.
The long-distance contributions modeled by triangle diagram at hadron level include
\begin{align}\label{x4}
L(T)_1[u,s,d,u]\,\,&=\,\, \frac{1}{2}\Delta(D^0,\pi^+,K^-,\rho^0,\pi^+,K^-)
-\frac{1}{2}\Delta(D^0,\pi^+,K^-,\omega,\pi^+,K^-),\\
L(E)[u,d,u,s]\,\,&=\,\, \frac{1}{2}\Delta(D^0,\pi^+,K^-,\rho^0,\pi^+,K^-)
+\frac{1}{2}\Delta(D^0,\pi^+,K^-,\omega,\pi^+,K^-).
\end{align}
After summing the $T$ and $E$ amplitudes, the rescattering amplitude of $D^0\to K^-\pi^+$ is
\begin{align}\label{x1}
\mathcal{A}_L(D^0\to K^-\pi^+)&= L(T)_1[u,s,d,u]+L(E)[u,d,u,s] = \Delta(D^0,\pi^+,K^-,\rho^0,\pi^+,K^-).
\end{align}
Notice that the contributions associated with $\omega\pi\pi$ vertex cancel each other.
The topological amplitude of $D^0\to \overline K^0\pi^0$ decay is $(C-E)/\sqrt{2}$.
The rescattering contributions include
\begin{align}
\frac{1}{\sqrt{2}}L(C)_1[u,u,d,s]\,\,&=\,\, -\frac{1}{\sqrt{2}}\Delta(D^0,\pi^+,K^-,\rho^+,\pi^0,\overline K^0),\\
\frac{1}{\sqrt{2}}L(C)_2[u,u,d,s]\,\,&=\,\, -\frac{1}{\sqrt{2}}\Delta(D^0,\pi^+,K^-,K^{*+},\overline K^0,\pi^0),\\
-\frac{1}{\sqrt{2}}L(E)[u,d,d,s]\,\,&=\,\, -\frac{1}{\sqrt{2}}\Delta(D^0,\pi^+,K^-,\rho^+,\pi^0,\overline K^0).
\end{align}
Summing the $C$ and $E$ amplitudes, we have
\begin{align}\label{x2}
\mathcal{A}_L(D^0\to \overline K^0\pi^0)=&\frac{1}{\sqrt{2}}L(C)_1[u,u,d,s]+\frac{1}{\sqrt{2}}L(C)_2[u,u,d,s]
-\frac{1}{\sqrt{2}}L(E)[u,d,d,s]\nonumber\\
=& -\sqrt{2}\Delta(D^0,\pi^+,K^-,\rho^+,\pi^0,\overline K^0)-\frac{1}{\sqrt{2}}\Delta(D^0,\pi^+,K^-,K^{*+},\overline K^0,\pi^0).
\end{align}
The topological amplitude of $D^+\to \overline K^0\pi^+$ decay is $T+C$.
The rescattering contributions include
\begin{align}\label{x5}
L(T)_2[d,s,d,u]\,\,&=\,\, -\frac{1}{2}\Delta(D^+,\pi^+,\overline K^0,\rho^0,\pi^+,\overline K^0)+\frac{1}{2}\Delta(D^0,\pi^+,\overline K^0,\omega,\pi^+,\overline K^0),\\
L(C)_1[d,u,d,s]\,\,&=\,\, -\frac{1}{2}\Delta(D^+,\pi^+,\overline K^0,\rho^0,\pi^+,\overline K^0)-\frac{1}{2}\Delta(D^+,\pi^+,\overline K^0,\omega,\pi^+,\overline K^0),\\
L(C)_2[d,u,d,s]\,\,&=\,\, \Delta(D^+,\pi^+,\overline K^0,K^{*+},\overline K^0,\pi^+).
\end{align}
Summing the $T$ and $C$ amplitudes, we have
\begin{align}\label{x3}
\mathcal{A}_L(D^+\to \overline K^0\pi^+)=&L(T)_2[d,s,d,u]+L(C)_1[d,u,d,s]+L(C)_2[d,u,d,s]\nonumber\\
=& -\Delta(D^+,\pi^+,\overline K^0,\rho^0,\pi^+,\overline K^0)+\Delta(D^+,\pi^+,\overline K^0,K^{*+},\overline K^0,\pi^+).
\end{align}
Again, all the contributions associated with $\omega\pi\pi$ vertex cancel each other.

Eqs.~\eqref{x1}, \eqref{x2} and \eqref{x3} can also be derived directly from the chiral Lagrangian~\eqref{vpp}.
That is, drawing and calculating the triangle diagram in a Feynman diagram like way with the chiral Lagrangian.
Starting from the topological diagrams, we obtain the same results via the topological-scattering diagrams.
It indicates the inherent correlation between the topological amplitude and the rescattering dynamics.
It is understandable since both the topological amplitude and rescattering dynamics are originated from the Quark Model (QM).

Under the Isospin symmetry, the particles in an Isospin multiplet can be seen as the same particles.
Then we have
\begin{align}
\Delta_1&= \Delta(D^0,\pi^+,K^-,\rho^0,\pi^+,K^-) = \Delta(D^0,\pi^+,K^-,\rho^+,\pi^0,\overline K^0) = \Delta(D^+,\pi^+,\overline K^0,\rho^0,\pi^+,\overline K^0), \nonumber\\
  \Delta_2&=\Delta(D^0,\pi^+,K^-,K^{*+},\overline K^0,\pi^0) = \Delta(D^+,\pi^+,\overline K^0,K^{*+},\overline K^0,\pi^+).
\end{align}
The decay amplitudes of $D^+\to \overline K^0\pi^+$, $D^0\to \overline K^0\pi^0$ and $D^0\to K^-\pi^+$ channels can be written as
\begin{align}
\mathcal{A}_L(D^+\to \overline K^0\pi^+)& =-\Delta_1-\Delta_2,\qquad \mathcal{A}_L(D^0\to \overline K^0\pi^0)=-\sqrt{2}\Delta_1-\frac{1}{\sqrt{2}}\Delta_2, \nonumber\\
\mathcal{A}_L(D^0\to K^-\pi^+)&=\Delta_1.
\end{align}
One can check the Isospin relation Eq.~\eqref{kpi} is satisfied in terms of the triangle diagrams.

The rescattering contributions of $T$ amplitude in the $D^0\to K^-\pi^+$ and $D^+\to \overline K^0\pi^+$ decays are different under the Isospin symmetry, see Eq.~\eqref{x4} and Eq.~\eqref{x5}.
In a higher flavor symmetry, the $SU(3)_F$ symmetry, the equivalence of $T$ diagram in the $D^0\to K^-\pi^+$ and $D^+\to \overline K^0\pi^+$ decays regained due to $\rho=\omega$.
Above discussion also applies to $C$ and $E$ diagrams.
It means the relation $L(C)=-2\,L(E)$ only holds in the $SU(3)_F$ symmetry but not in the Isospin symmetry.
The Isospin amplitudes of the $D^0\to K^-\pi^+$, $D^0\to \overline K^0\pi^0$ and $D^+\to \overline K^0\pi^+$ decays are
\begin{align}
\mathcal{A}(D^+\to \overline K^0\pi^+)& =\frac{1}{3}\mathcal{A}_{3/2}+\frac{2}{3}\mathcal{A}_{1/2},\qquad
 \mathcal{A}(D^0\to \overline K^0\pi^0)=\frac{\sqrt{2}}{3}(\mathcal{A}_{3/2}-\mathcal{A}_{1/2}), \nonumber\\
\mathcal{A}(D^0\to K^-\pi^+)&=\mathcal{A}_{3/2}.
\end{align}
Notice that there are only two Isospin amplitudes.  The number of Isospin amplitude does not equal to the topological amplitude.
Thus not all topological amplitudes are physical in the Isospin symmetry.
The topological amplitudes $T$, $C$ and $E$ need not be equivalent respectively in different decay channels if only the Isospin symmetry is maintained.

The decay modes $D^0\to K^0\pi^0$, $D^0\to K^+\pi^-$, $D^+\to K^+\pi^0$ and $D^+\to K^0\pi^+$ form another Isospin relation for $D\to K\pi$ decays \cite{Grossman:2012ry}:
\begin{align}\label{kpi2}
\sqrt{2} \mathcal{A}(D^0\to K^0\pi^0)+\mathcal{A}(D^0\to K^+\pi^-)+\sqrt{2}\mathcal{A}(D^+\to K^+\pi^0)-\mathcal{A}(D^+\to K^0\pi^+)=0.
\end{align}
The topological amplitude of $D^0\to K^0\pi^0$ decay is $(C-E)/\sqrt{2}$.
The long-distance contributions include
\begin{align}
\frac{1}{\sqrt{2}}L(C)_1[u,u,s,d]\,\,&=\,\, -\frac{1}{\sqrt{2}}\Delta(D^0,K^+,\pi^-,K^{*+},\pi^0, K^0),\\
\frac{1}{\sqrt{2}}L(C)_2[u,u,s,d]\,\,&=\,\, -\frac{1}{\sqrt{2}}\Delta(D^0,K^+,\pi^-,\rho^+,K^0, \pi^0),\\
-\frac{1}{\sqrt{2}}L(E)[u,s,d,d]\,\,&=\,\, -\frac{1}{\sqrt{2}}\Delta(D^0,K^+,\pi^-,\rho^+,K^0, \pi^0).
\end{align}
Summing them, the rescattering amplitude of $D^0\to K^0\pi^0$ is
\begin{align}
\mathcal{A}_L(D^0\to K^0\pi^0)&= \frac{1}{\sqrt{2}}L(C)_1[u,u,s,d]+\frac{1}{\sqrt{2}}L(C)_2[u,u,s,d]-\frac{1}{\sqrt{2}}L(E)[u,s,d,d]\nonumber\\&= -\sqrt{2}\Delta(D^0,K^+,\pi^-,\rho^+,K^0, \pi^0)-\frac{1}{\sqrt{2}}\Delta(D^0,K^+,\pi^-,K^{*+},\pi^0, K^0).
\end{align}
The topological amplitude of $D^0\to K^+\pi^-$ decay is $E+T$.
The long-distance contributions include
\begin{align}
L(E)[u,s,u,d]\,\,&=\,\, \frac{1}{2}\Delta(D^0,K^+,\pi^-,\rho^0,K^+, \pi^-)+\frac{1}{2}\Delta(D^0,K^+,\pi^-,\omega^0,K^+,\pi^-),\\
L(T)_4[u,d,s,u]\,\,&=\,\, \frac{1}{2}\Delta(D^0,K^+,\pi^-,\rho^0,K^+, \pi^-)-\frac{1}{2}\Delta(D^0,K^+,\pi^-,\omega^0,K^+,\pi^-).
\end{align}
Summing them, the rescattering amplitude of $D^0\to K^+\pi^-$ is
\begin{align}
\mathcal{A}_L(D^0\to K^+\pi^-)&= L(E)[u,s,u,d]+L(T)_4[u,d,s,u]=\Delta(D^0,K^+,\pi^-,\rho^0,K^+, \pi^-).
\end{align}

The topological amplitude of $D^+\to K^+\pi^0$ decay is $(A-T)/\sqrt{2}$.
The long-distance contributions include
\begin{align}
\frac{1}{\sqrt{2}}L(A)_1[d,s,u,u]\,\,=& \,\, \frac{1}{4\sqrt{2}}\Delta(D^+,K^+,\pi^0,\rho^0,K^+, \pi^0)-\frac{1}{12\sqrt{2}}\Delta(D^+,K^+,\eta_8,\rho^0,K^+, \pi^0)\nonumber\\&-\frac{1}{6\sqrt{2}}\Delta(D^+,K^+,\eta_1,\rho^0,K^+, \pi^0)+\frac{1}{4\sqrt{2}}\Delta(D^+,K^+,\pi^0,\omega,K^+, \pi^0)\nonumber\\& -\frac{1}{12\sqrt{2}}\Delta(D^+,K^+,\eta_8,\omega,K^+, \pi^0)-\frac{1}{6\sqrt{2}}\Delta(D^+,K^+,\eta_1,\omega,K^+, \pi^0),\\
\frac{1}{\sqrt{2}}L(A)_2[d,s,u,u]\,\,=& \,\, \frac{1}{2\sqrt{2}}\Delta(D^+,K^+,\pi^0,K^{*+},\pi^0,K^+)-\frac{1}{6\sqrt{2}}\Delta(D^+,K^+,\eta_8,K^{*+},\pi^0,K^+)\nonumber\\&
-\frac{1}{3\sqrt{2}}\Delta(D^+,K^+,\eta_1,K^{*+},\pi^0,K^+),\\
\frac{1}{\sqrt{2}}L(A)_3[d,s,u,u]\,\,=& \,\, -\frac{1}{4\sqrt{2}}\Delta(D^+,K^+,\pi^0,\rho^0,K^+, \pi^0)+\frac{1}{12\sqrt{2}}\Delta(D^+,K^+,\eta_8,\rho^0,K^+, \pi^0)\nonumber\\&+\frac{1}{6\sqrt{2}}\Delta(D^+,K^+,\eta_1,\rho^0,K^+, \pi^0)-\frac{1}{4\sqrt{2}}\Delta(D^+,K^+,\pi^0,\omega,K^+, \pi^0)\nonumber\\& +\frac{1}{12\sqrt{2}}\Delta(D^+,K^+,\eta_8,\omega,K^+, \pi^0)+\frac{1}{6\sqrt{2}}\Delta(D^+,K^+,\eta_1,\omega,K^+, \pi^0),\\
\frac{1}{\sqrt{2}}L(A)_4[d,s,u,u]\,\,=& \,\, -\frac{1}{3\sqrt{2}}\Delta(D^+,K^+,\eta_8,K^{*+},\pi^0,K^+)
+\frac{1}{3\sqrt{2}}\Delta(D^+,K^+,\eta_1,K^{*+},\pi^0,K^+),\\
-\frac{1}{\sqrt{2}}L(T)_2[d,d,s,u]\,\,=& \,\, \frac{1}{4\sqrt{2}}\Delta(D^+,K^+,\pi^0,\rho^0,K^+, \pi^0)+\frac{1}{12\sqrt{2}}\Delta(D^+,K^+,\eta_8,\rho^0,K^+, \pi^0)\nonumber\\&+\frac{1}{6\sqrt{2}}\Delta(D^+,K^+,\eta_1,\rho^0,K^+, \pi^0)-\frac{1}{4\sqrt{2}}\Delta(D^+,K^+,\pi^0,\omega,K^+, \pi^0)\nonumber\\& -\frac{1}{12\sqrt{2}}\Delta(D^+,K^+,\eta_8,\omega,K^+, \pi^0)-\frac{1}{6\sqrt{2}}\Delta(D^+,K^+,\eta_1,\omega,K^+, \pi^0),\\
-\frac{1}{\sqrt{2}}L(T)_4[d,d,s,u]\,\,=& \,\, -\frac{1}{4\sqrt{2}}\Delta(D^+,K^+,\pi^0,\rho^0,K^+, \pi^0)-\frac{1}{12\sqrt{2}}\Delta(D^+,K^+,\eta_8,\rho^0,K^+, \pi^0)\nonumber\\&-\frac{1}{6\sqrt{2}}\Delta(D^+,K^+,\eta_1,\rho^0,K^+, \pi^0)+\frac{1}{4\sqrt{2}}\Delta(D^+,K^+,\pi^0,\omega,K^+, \pi^0)\nonumber\\& +\frac{1}{12\sqrt{2}}\Delta(D^+,K^+,\eta_8,\omega,K^+, \pi^0)+\frac{1}{6\sqrt{2}}\Delta(D^+,K^+,\eta_1,\omega,K^+, \pi^0).
\end{align}
Summing them, the rescattering amplitude of $D^+\to K^+\pi^0$ is
\begin{align}
\mathcal{A}_L(D^+\to K^+\pi^0)& = \frac{1}{\sqrt{2}}(L(A)_1[d,s,u,u]+L(A)_2[d,s,u,u]+L(A)_3[d,s,u,u]+L(A)_4[d,s,u,u])\nonumber\\&~~~-\frac{1}{\sqrt{2}}(L(T)_2[d,d,s,u]+L(T)_4[u,d,s,u])
\nonumber\\& = \frac{1}{2\sqrt{2}}\Delta(D^+,K^+,\pi^0,K^{*+},\pi^0,K^+)-\frac{1}{2\sqrt{2}}\Delta(D^+,K^+,\eta_8,K^{*+},\pi^0,K^+).
\end{align}
The topological amplitude of $D^+\to K^0\pi^+$ decay is $C+A$.
The long-distance contributions include
\begin{align}
L(C)_1[d,u,s,d]\,\,=&\,\, -\frac{1}{2}\Delta(D^+,K^+,\pi^0,K^{*0},\pi^+,K^0)-\frac{1}{6}\Delta(D^+,K^+,\eta_8,K^{*0},\pi^+,K^0)\nonumber\\&
-\frac{1}{3}\Delta(D^+,K^+,\eta_1,K^{*0},\pi^+,K^0),\\
L(C)_2[d,u,s,d]\,\,&=\,\,  -\frac{1}{2}\Delta(D^+,K^+,\pi^0,\rho^+,K^0,\pi^+)-\frac{1}{6}\Delta(D^+,K^+,\eta_8,\rho^+,K^0,\pi^+)\nonumber\\&
-\frac{1}{3}\Delta(D^+,K^+,\eta_1,\rho^+,K^0,\pi^+),\\
L(A)_3[d,s,d,u]\,\,&=\,\,  -\frac{1}{2}\Delta(D^+,K^+,\pi^0,\rho^+,K^0,\pi^+)+\frac{1}{6}\Delta(D^+,K^+,\eta_8,\rho^+,K^0,\pi^+)\nonumber\\&
+\frac{1}{3}\Delta(D^+,K^+,\eta_1,\rho^+,K^0,\pi^+),\\
L(A)_4[d,s,d,u]\,\,=&\,\,  -\frac{1}{3}\Delta(D^+,K^+,\eta_8,K^{*0},\pi^+,K^0)+\frac{1}{3}\Delta(D^+,K^+,\eta_1,K^{*0},\pi^+,K^0).
\end{align}
Summing them, the rescattering amplitude of $D^+\to K^0\pi^+$ is
\begin{align}
\mathcal{A}_L(D^+\to K^0\pi^+) = &L(C)_1[d,u,s,d]+L(C)_2[d,u,s,d]+L(A)_3[d,s,d,u]+L(A)_4[d,s,d,u]\nonumber\\ =&-\Delta(D^+,K^+,\pi^0,\rho^+,K^0,\pi^+)
-\frac{1}{2}\Delta(D^+,K^+,\pi^0,K^{*0},\pi^+,K^0)\nonumber\\&-\frac{1}{2}\Delta(D^+,K^+,\eta_8,K^{*0},\pi^+,K^0).
\end{align}
Under the Isospin symmetry, we have
\begin{align}
\Delta_3&= \Delta(D^0,K^+,\pi^-,\rho^+,K^0, \pi^0) = \Delta(D^0,K^+,\pi^-,\rho^0,K^+, \pi^-)=\Delta(D^+,K^+,\pi^0,K^{*0},\pi^+,K^0), \nonumber\\
\Delta_4&= \Delta(D^0,K^+,\pi^-,K^{*+},\pi^0, K^0) = \Delta(D^+,K^+,\pi^0,K^{*+},\pi^0,K^+)=\Delta(D^+,K^+,\pi^0,K^{*0},\pi^+,K^0), \nonumber\\
\Delta_5&= \Delta(D^+,K^+,\eta_8,K^{*+},\pi^0,K^+)=\Delta(D^+,K^+,\eta_8,K^{*0},\pi^+,K^0).
\end{align}
The decay amplitudes of $D^0\to K^0\pi^0$, $D^0\to K^+\pi^-$, $D^+\to K^+\pi^0$ and $D^+\to K^0\pi^+$ channels can be written as
\begin{align}
\mathcal{A}_L(D^0\to K^0\pi^0)&= -\sqrt{2}\Delta_3-\frac{1}{\sqrt{2}}\Delta_4, \qquad \mathcal{A}_L(D^0\to K^+\pi^-)=\Delta_3,\nonumber\\
\mathcal{A}_L(D^+\to K^+\pi^0)& = \frac{1}{2\sqrt{2}}\Delta_4-\frac{1}{2\sqrt{2}}\Delta_5,  \qquad \mathcal{A}_L(D^+\to K^0\pi^+) =- \Delta_3
-\frac{1}{2}\Delta_4-\frac{1}{2}\Delta_5.
\end{align}
One can check the Isospin relation Eq.~\eqref{kpi2} is satisfied.

\subsection{$D\to \pi\pi$ decay}\label{dpipi}
The decay modes $D^0\to \pi^+\pi^-$, $D^0\to \pi^0\pi^0$ and $D^+\to \pi^+\pi^0$ have following Isospin relation
\begin{align}\label{pipi}
 \mathcal{A}(D^0\to \pi^+\pi^-)-\sqrt{2}\mathcal{A}(D^0\to \pi^0\pi^0)+\sqrt{2}\mathcal{A}(D^+\to \pi^+\pi^0)=0.
\end{align}
The topological amplitude of $D^0\to \pi^+\pi^-$ decay is $\lambda_d(T+E)+\lambda_d(P+2PA)+\lambda_s(P+2PA)$, in which $\lambda_d=V^*_{cd}V_{ud}$ and $\lambda_s=V^*_{cs}V_{us}$.
The rescattering contributions in the $D^0\to \pi^+\pi^-$ decay modeled by triangle diagram include
\begin{align}
\lambda_dL(T)_1[u,d,d,u]\,\,&=\,\, \frac{1}{2}\lambda_d\Delta(D^0,\pi^+,\pi^-,\rho^0,\pi^+,\pi^-)
-\frac{1}{2}\lambda_d\Delta(D^0,\pi^+,\pi^-,\omega,\pi^+,\pi^-),\\
\lambda_dL(T)_4[u,d,d,u]\,\,&=\,\, \frac{1}{2}\lambda_d\Delta(D^0,\pi^+,\pi^-,\rho^0,\pi^+,\pi^-)
-\frac{1}{2}\lambda_d\Delta(D^0,\pi^+,\pi^-,\omega,\pi^+,\pi^-),\\
\lambda_dL(E)[u,d,u,d]\,\,&=\,\, \frac{1}{2}\lambda_d\Delta(D^0,\pi^+,\pi^-,\rho^0,\pi^+,\pi^-)
+\frac{1}{2}\lambda_d\Delta(D^0,\pi^+,\pi^-,\omega,\pi^+,\pi^-),\\
\lambda_dL(P)[u,d,u,d]\,\,&=\,\, \frac{1}{2}\lambda_d\Delta(D^0,\pi^+,\pi^-,\rho^0,\pi^+,\pi^-)
+\frac{1}{2}\lambda_d\Delta(D^0,\pi^+,\pi^-,\omega,\pi^+,\pi^-),\\
\lambda_sL(P)[u,d,u,s]\,\,&=\,\, \lambda_s\Delta(D^0,K^+,K^-,K^{*0},\pi^+,\pi^-).
\end{align}
Summing all of them, the rescattering amplitude in the $D^0\to \pi^+\pi^-$ decay is
\begin{align}\label{x6}
\mathcal{A}_L(D^0\to \pi^+\pi^-)=&\lambda_d(L(T)_1[u,d,d,u]+L(T)_4[u,d,d,u]+L(E)[u,d,u,d]\nonumber\\&
+L(P)[u,d,u,d])+\lambda_sL(P)[u,d,u,s]\nonumber\\
=& 2\lambda_d\Delta(D^0,\pi^+,\pi^-,\rho^0,\pi^+,\pi^-)
+\lambda_s\Delta(D^0,K^+,K^-,K^{*0},\pi^+,\pi^-).
\end{align}
The topological amplitude of $D^0\to \pi^0\pi^0$ decay is $\lambda_d(E-C)/\sqrt{2}+\lambda_d(P+2PA)/\sqrt{2}
+\lambda_s(P+2PA)/\sqrt{2}$.
The rescattering contributions in the $D^0\to \pi^0\pi^0$ decay include
\begin{align}
\frac{1}{\sqrt{2}}\lambda_dL(E)[u,d,d,d]\,\,&=\,\, \frac{1}{\sqrt{2}}\lambda_d\Delta(D^0,\pi^+,\pi^-,\rho^+,\pi^0,\pi^0),\\
-\frac{1}{\sqrt{2}}\lambda_dL(C)_1[u,u,d,d]\,\,&=\,\, \frac{1}{\sqrt{2}}\lambda_d\Delta(D^0,\pi^+,\pi^-,\rho^+,\pi^0,\pi^0),\\
-\frac{1}{\sqrt{2}}\lambda_dL(C)_2[u,u,d,d]\,\,&=\,\, \frac{1}{\sqrt{2}}\lambda_d\Delta(D^0,\pi^+,\pi^-,\rho^+,\pi^0,\pi^0),\\
\frac{1}{\sqrt{2}}\lambda_dL(P)[u,u,u,d]\,\,&=\,\, \frac{1}{\sqrt{2}}\lambda_d\Delta(D^0,\pi^+,\pi^-,\rho^+,\pi^0,\pi^0),\\
\frac{1}{\sqrt{2}}\lambda_sL(P)[u,u,u,s]\,\,&=\,\, \frac{1}{\sqrt{2}}\lambda_s\Delta(D^0,K^+,K^-,K^{*+},\pi^0,\pi^0).
\end{align}
Summing all of them, the rescattering amplitude in the $D^0\to \pi^0\pi^0$ decay is
\begin{align}\label{x7}
\mathcal{A}_L(D^0\to \pi^0\pi^0)=&\frac{1}{\sqrt{2}}\lambda_d(L(E)[u,d,d,d]-L(C)_1[u,u,d,d]
-L(C)_2[u,u,d,d]\nonumber\\&+L(P)[u,u,u,d])
+\frac{1}{\sqrt{2}}\lambda_sL(P)[u,u,u,s]\nonumber\\
=& 2\sqrt{2}\lambda_d\Delta(D^0,\pi^+,\pi^-,\rho^+,\pi^0,\pi^0)
+\frac{1}{\sqrt{2}}\lambda_s\Delta(D^0,K^+,K^-,K^{*+},\pi^0,\pi^0).
\end{align}
The topological amplitude of $D^+\to \pi^+\pi^0$ decay is $-\lambda_d(T+C)/\sqrt{2}$.
The rescattering contributions in the $D^+\to \pi^+\pi^0$ decay include
\begin{align}
-\frac{1}{\sqrt{2}}\lambda_dL(T)_2[d,d,u,d]\,\,=&\,\, \frac{1}{4\sqrt{2}}\lambda_d\Delta(D^+,\pi^+,\pi^0,\rho^0,\pi^+,\pi^0)
+\frac{1}{12\sqrt{2}}\lambda_d\Delta(D^0,\pi^+, \eta_8,\rho^0,\pi^+,\pi^0)\nonumber\\ & +\frac{1}{6\sqrt{2}}\lambda_d\Delta(D^+,\pi^+,\eta_1,\rho^0,\pi^+,\pi^0)
-\frac{1}{4\sqrt{2}}\lambda_d\Delta(D^+,\pi^+,\pi^0,\omega,\pi^+,\pi^0)\nonumber\\ & -\frac{1}{12\sqrt{2}}\lambda_d\Delta(D^0,\pi^+, \eta_8,\omega,\pi^+,\pi^0)
-\frac{1}{6\sqrt{2}}\lambda_d\Delta(D^+,\pi^+,\eta_1,\omega,\pi^+,\pi^0),\\
-\frac{1}{\sqrt{2}}\lambda_dL(T)_4[d,d,u,d]\,\,=&\,\, -\frac{1}{4\sqrt{2}}\lambda_d\Delta(D^+,\pi^+,\pi^0,\rho^0,\pi^+,\pi^0)
-\frac{1}{12\sqrt{2}}\lambda_d\Delta(D^0,\pi^+, \eta_8,\rho^0,\pi^+,\pi^0)\nonumber\\ &=\,\, -\frac{1}{6\sqrt{2}}\lambda_d\Delta(D^+,\pi^+,\eta_1,\rho^0,\pi^+,\pi^0)
+\frac{1}{4\sqrt{2}}\lambda_d\Delta(D^+,\pi^+,\pi^0,\omega,\pi^+,\pi^0)\nonumber\\ &+\frac{1}{12\sqrt{2}}\lambda_d\Delta(D^0,\pi^+, \eta_8,\omega,\pi^+,\pi^0)
+\frac{1}{6\sqrt{2}}\lambda_d\Delta(D^+,\pi^+,\eta_1,\omega,\pi^+,\pi^0),\\
-\frac{1}{\sqrt{2}}\lambda_dL(C)_1[d,u,d,d]\,\,=&\,\, \frac{1}{4\sqrt{2}}\lambda_d\Delta(D^+,\pi^+,\pi^0,\rho^0,\pi^+,\pi^0)
+\frac{1}{12\sqrt{2}}\lambda_d\Delta(D^0,\pi^+, \eta_8,\rho^0,\pi^+,\pi^0)\nonumber\\ & +\frac{1}{6\sqrt{2}}\lambda_d\Delta(D^+,\pi^+,\eta_1,\rho^0,\pi^+,\pi^0)
+\frac{1}{4\sqrt{2}}\lambda_d\Delta(D^+,\pi^+,\pi^0,\omega,\pi^+,\pi^0)\nonumber\\ & +\frac{1}{12\sqrt{2}}\lambda_d\Delta(D^0,\pi^+, \eta_8,\omega,\pi^+,\pi^0)
+\frac{1}{6\sqrt{2}}\lambda_d\Delta(D^+,\pi^+,\eta_1,\omega,\pi^+,\pi^0),\\
-\frac{1}{\sqrt{2}}\lambda_dL(C)_2[d,u,d,d]\,\,=&\,\, \frac{1}{2\sqrt{2}}\lambda_d\Delta(D^+,\pi^+,\pi^0,\rho^+,\pi^0,\pi^+)
+\frac{1}{6\sqrt{2}}\lambda_d\Delta(D^0,\pi^+, \eta_8,\rho^+,\pi^0,\pi^+)\nonumber\\ & +\frac{1}{3\sqrt{2}}\lambda_d\Delta(D^+,\pi^+,\eta_1,\rho^+,\pi^0,\pi^+).
\end{align}
Topology $A$ does not contribute to the $D^+\to \pi^+\pi^0$ decay in the $SU(3)_F$ symmetry because of the cancellation between $u\bar u$ and $d\bar d$ generated via gluons.
In order to get completed rescattering contributions, the topological-scattering diagrams in the $T\Rightarrow A$ transition should be considered.
The rescattering contributions in $A$ diagram are
\begin{align}
\frac{1}{\sqrt{2}}\lambda_dL(A)_1[d,d,u,u]\,\,=&\,\, \frac{1}{4\sqrt{2}}\lambda_d\Delta(D^+,\pi^+,\pi^0,\rho^0,\pi^+,\pi^0)
-\frac{1}{12\sqrt{2}}\lambda_d\Delta(D^0,\pi^+, \eta_8,\rho^0,\pi^+,\pi^0)\nonumber\\ & -\frac{1}{6\sqrt{2}}\lambda_d\Delta(D^+,\pi^+,\eta_1,\rho^0,\pi^+,\pi^0)
+\frac{1}{4\sqrt{2}}\lambda_d\Delta(D^+,\pi^+,\pi^0,\omega,\pi^+,\pi^0)\nonumber\\ & -\frac{1}{12\sqrt{2}}\lambda_d\Delta(D^0,\pi^+, \eta_8,\omega,\pi^+,\pi^0)
-\frac{1}{6\sqrt{2}}\lambda_d\Delta(D^+,\pi^+,\eta_1,\omega,\pi^+,\pi^0),\\
\frac{1}{\sqrt{2}}\lambda_dL(A)_2[d,d,u,u]\,\,=&\,\, \frac{1}{2\sqrt{2}}\lambda_d\Delta(D^+,\pi^+,\pi^0,\rho^+,\pi^0,\pi^+)
-\frac{1}{6\sqrt{2}}\lambda_d\Delta(D^0,\pi^+, \eta_8,\rho^+,\pi^0,\pi^+)\nonumber\\ & -\frac{1}{3\sqrt{2}}\lambda_d\Delta(D^+,\pi^+,\eta_1,\rho^+,\pi^0,\pi^+),\\
\frac{1}{\sqrt{2}}\lambda_dL(A)_3[d,d,u,u]\,\,=&\,\, -\frac{1}{4\sqrt{2}}\lambda_d\Delta(D^+,\pi^+,\pi^0,\rho^0,\pi^+,\pi^0)
+\frac{1}{12\sqrt{2}}\lambda_d\Delta(D^0,\pi^+, \eta_8,\rho^0,\pi^+,\pi^0)\nonumber\\ & +\frac{1}{6\sqrt{2}}\lambda_d\Delta(D^+,\pi^+,\eta_1,\rho^0,\pi^+,\pi^0)
-\frac{1}{4\sqrt{2}}\lambda_d\Delta(D^+,\pi^+,\pi^0,\omega,\pi^+,\pi^0)\nonumber\\ & +\frac{1}{12\sqrt{2}}\lambda_d\Delta(D^0,\pi^+, \eta_8,\omega,\pi^+,\pi^0)
+\frac{1}{6\sqrt{2}}\lambda_d\Delta(D^+,\pi^+,\eta_1,\omega,\pi^+,\pi^0),\\
-\frac{1}{\sqrt{2}}\lambda_dL(A)_3[d,d,d,u]\,\,=&\,\, \frac{1}{2\sqrt{2}}\lambda_d\Delta(D^+,\pi^+,\pi^0,\rho^+,\pi^0,\pi^+)
-\frac{1}{6\sqrt{2}}\lambda_d\Delta(D^0,\pi^+, \eta_8,\rho^+,\pi^0,\pi^+)\nonumber\\ & -\frac{1}{3\sqrt{2}}\lambda_d\Delta(D^+,\pi^+,\eta_1,\rho^+,\pi^0,\pi^+).
\end{align}
Similar to $A$ diagram, the rescattering contributions in $P$ diagram should also be considered:
\begin{align}
\frac{1}{\sqrt{2}}\lambda_dL(P)[d,u,u,d]\,\,=&\,\, \frac{1}{2\sqrt{2}}\lambda_d\Delta(D^+,\pi^+,\pi^0,\rho^+,\pi^0,\pi^+)
+\frac{1}{6\sqrt{2}}\lambda_d\Delta(D^0,\pi^+, \eta_8,\rho^+,\pi^0,\pi^+)\nonumber\\ & +\frac{1}{3\sqrt{2}}\lambda_d\Delta(D^+,\pi^+,\eta_1,\rho^+,\pi^0,\pi^+),\\
-\frac{1}{\sqrt{2}}\lambda_dL(P)[d,d,u,d]\,\,=&\,\, -\frac{1}{4\sqrt{2}}\lambda_d\Delta(D^+,\pi^+,\pi^0,\rho^0,\pi^+,\pi^0)
-\frac{1}{12\sqrt{2}}\lambda_d\Delta(D^0,\pi^+, \eta_8,\rho^0,\pi^+,\pi^0)\nonumber\\ & -\frac{1}{6\sqrt{2}}\lambda_d\Delta(D^+,\pi^+,\eta_1,\rho^0,\pi^+,\pi^0)
-\frac{1}{4\sqrt{2}}\lambda_d\Delta(D^+,\pi^+,\pi^0,\omega,\pi^+,\pi^0)\nonumber\\ & -\frac{1}{12\sqrt{2}}\lambda_d\Delta(D^0,\pi^+, \eta_8,\omega,\pi^+,\pi^0)
-\frac{1}{6\sqrt{2}}\lambda_d\Delta(D^+,\pi^+,\eta_1,\omega,\pi^+,\pi^0),\\
\frac{1}{\sqrt{2}}\lambda_sL(P)[d,u,u,s]\,\,&=\,\, \frac{1}{\sqrt{2}}\lambda_s\Delta(D^0,K^+,\overline K^0,K^{*+},\pi^0,\pi^+),\\
-\frac{1}{\sqrt{2}}\lambda_sL(P)[d,d,u,s]\,\,&=\,\, -\frac{1}{\sqrt{2}}\lambda_s\Delta(D^0,K^+,\overline K^0, K^{*0},\pi^+,\pi^0).
\end{align}
Summing all topological-scattering diagrams, the rescattering amplitude in the $D^+\to \pi^+\pi^0$ decay is
\begin{align}\label{x8}
\mathcal{A}_L(D^+\to \pi^+\pi^0)=&-\frac{1}{\sqrt{2}}\lambda_d(L(T)_2[d,d,u,d]+L(T)_4[d,d,u,d]
+L(C)_1[d,u,d,d]\nonumber\\&
+L(C)_2[d,u,d,d]-L(A)_1[d,d,u,u]-L(A)_2[d,d,u,u]\nonumber\\&-L(A)_3[d,d,u,u]
+L(A)_3[d,d,d,u]
-L(P)[d,u,u,d]\nonumber\\&+L(P)[d,d,u,d])
+\frac{1}{\sqrt{2}}\lambda_s(L(P)[d,u,u,s]-L(P)[d,d,u,s])
\nonumber\\=& \sqrt{2}\lambda_d\Delta(D^0,\pi^+,\pi^0,\rho^+,\pi^+,\pi^0)
+\frac{1}{\sqrt{2}}\lambda_s(\Delta(D^0,K^+,\overline K^0,K^{*+},\pi^0,\pi^+)\nonumber\\&
-\Delta(D^0,K^+,\overline K^0, K^{*0},\pi^+,\pi^0)).
\end{align}
Notice that all the contributions with vertexes that do not appear in the chiral Lagrangian~\eqref{vpp}, such as $\rho^0\pi^0\pi^0$, $\rho^0\eta\eta$ ..., cancel each other.

Eqs.~\eqref{x6}, \eqref{x7} and \eqref{x8} are consistent with the rescattering amplitudes derived directly from the chiral Lagrangian.
Under the Isospin symmetry, we have
\begin{align}
\Delta_6&= \Delta(D^0,\pi^+,\pi^-,\rho^0,\pi^+,\pi^-) = \Delta(D^0,\pi^+,\pi^-,\rho^+,\pi^0,\pi^0) = \Delta(D^0,\pi^+,\pi^0,\rho^+,\pi^+,\pi^0), \nonumber\\
  \Delta_7&=\Delta(D^0,K^+,K^-,K^{*0},\pi^+,\pi^-) = \Delta(D^0,K^+,K^-,K^{*+},\pi^0,\pi^0)\nonumber\\&~~~~~~=\Delta(D^0,K^+,\overline K^0,K^{*+},\pi^0,\pi^+)=\Delta(D^0,K^+,\overline K^0, K^{*0},\pi^+,\pi^0).
\end{align}
Then the amplitudes of $D^0\to \pi^+\pi^-$, $D^0\to \pi^0\pi^0$ and $D^+\to \pi^+\pi^0$ channels can be written as
\begin{align}
\mathcal{A}_L(D^0\to \pi^+\pi^-)& =2\lambda_d\Delta_6+\lambda_s\Delta_7,\qquad \mathcal{A}_L(D^0\to \pi^0\pi^0)=2\sqrt{2}\lambda_d\Delta_6+\frac{1}{\sqrt{2}}\lambda_s\Delta_7, \nonumber\\
\mathcal{A}_L(D^+\to \pi^+\pi^0)&=\sqrt{2}\lambda_d\Delta_6.
\end{align}
One can check the Isospin relation Eq.~\eqref{pipi} is satisfied.

In \cite{Li:2002pj}, the Isospin factor is considered in such a way that the $u\bar u$ component in one final meson $\pi^0$ contributes a factor of $1/\sqrt{2}$ and the $d\bar d$ component contributes a factor of $-1/\sqrt{2}$.
For the intermediate state $\pi^0$, the factor $1/\sqrt{2}$ or $-1/\sqrt{2}$ is dropped to keep the Isospin relation between the $D^0\to \pi^+\pi^-$, $D^0\to \pi^0\pi^0$ and $D^+\to \pi^+\pi^0$ channels.
This operation is artificial without sufficient reason.
The confusion comes from that not all the topological-scattering diagrams are considered in \cite{Li:2002pj}.
In this work, the complete topological-scattering diagrams contributing to the three channels are found via the tensor form of topological-scattering diagram.
The Isospin factor is derived directly form topological-scattering diagram. For the neutral meson serving an intermediate state, the Isospin factor is multiplied two times because of the two vertexes. Summing all the topological-scattering diagrams, the Isospin relation holds in the amplitudes expressed in the triangle diagrams.

The framework proposed in this work can be extended to other decay modes, such as $B$ meson or heavy baryon decays. Taking the $\overline B\to D\pi$ and $\overline B\to \pi\pi$ decays as examples, we discuss the applications in the $B$ meson decays in Appendix.~\ref{bdecay}.
One can find the conclusions in the $D$ meson decays under the $SU(3)_F$ symmetry can be generalized into the $B$ meson decays under the $SU(4)_F$ symmetry although the $SU(4)_F$ symmetry is expected to be much less precise than the $SU(3)_F$ symmetry.
For the heavy baryon decays, we shall leave them to the future work.

\section{Summary}\label{sum}

In this work, we proposed a theoretical framework to correlate the topological diagram at quark level and rescattering dynamics at hadron level.
The main points are the following:
\begin{enumerate}
\item Both the topological diagram, topological-scattering diagram, and triangle diagram can be written in the tensor form.
\item The coefficient of each triangle diagram can be derived from the topological-scattering diagram.
\item There is one minus sign between the topological-scattering diagrams with and without a cross of quark lines. It arises from the commutator in the chiral Lagrangian.
\item The vertexes such as $\omega\pi^+\pi^-$, $\rho^0\pi^0\pi^0$ ... might appear in a triangle diagrams derived from topological-scattering diagrams.
    But all the contributions associated with them cancel after summing all triangle diagrams in one decay channel.
\item  There are no triangle diagram like long-distance contributions in the topologies $ES$, $AS$, $PS$, $PA$ and $SS$ because they can be divided into several unconnected substructures by cutting off gluon propagators.
\item The triangle diagrams derived from the topological-scattering diagrams are consistent with the ones derived from the chiral Lagrangian directly.
\item The completeness of the topological-scattering diagram is confirmed by the quark substructure of meson-meson scattering since there is no repeated topological-scattering diagrams in the $T \Rightarrow T$, $T \Rightarrow C$, $T \Rightarrow E$, $T \Rightarrow A$ and $T \Rightarrow P$ transitions and all the twelve possible substructures are included in the topological-scattering diagrams.
\item The rescattering contributions arisen from $C^{SD}$ can be obtained from the ones arisen from $T^{SD}$ by replacing $T^{SD}$ with $C^{SD}$ in the twelve topological-scattering diagrams.
\item Under the $SU(3)_F$ symmetry, the rescattering contribution arisen from $T^{SD}$ in the $C$ diagram is $-2$ times the ones in the $E$ and $P$ diagrams, $L(C):L(E):L(P)=-2:1:1$. It indicates the penguin amplitudes are the same order with the tree amplitudes, which leads to an observable CP violation in charm decay.
\item The rescattering contributions arisen from $T^{SD}$ in the $T$ and $A$ diagrams are arisen from the $SU(3)_F$ breaking effects.
    It could be understood in the large $N_c$ expansion since the meson-meson scattering in the topological-scattering diagrams of $T\Rightarrow T$ and $T\Rightarrow A$ transitions is at order of $1/N_c^2$ compared to $1/N_c$ in the $T\Rightarrow C$, $T\Rightarrow E$ and $T\Rightarrow P$ transitions. Considering that $C^{SD}$ is one order smaller than $T^{SD}$, the rescattering contributions in the $T$ and $A$ diagrams are smaller than $C$, $E$ and $P$ diagrams.
\item The Isospin relations in some decay channels hold in terms of the triangle diagrams.
\item The conclusions about $D$ meson decays under the $SU(3)_F$ symmetry are also valid in the $B$ meson decays under the $SU(4)_F$ symmetry.
\end{enumerate}

\begin{acknowledgements}
We are grateful to Hai-Yang Cheng, Fu-Sheng Yu and Cai-Ping Jia for useful discussions.
This work was supported in part by the National Natural Science Foundation of China
under Grants No.12105099.
\end{acknowledgements}

\begin{appendix}

\section{Application in $B$ meson decay}\label{bdecay}
In this Appendix, we analyze the $\overline B\to D\pi$ and $\overline B\to \pi\pi$ decays as supplementary examples.

\subsection{$\overline B\to D\pi$ decay}\label{bkpi}

The topological amplitude of $\overline B^0\to D^+\pi^-$ decay is $T+E$.
The long-distance contributions modeled by triangle diagram include
\begin{align}
L(T)_1[d,c,u,d]\,\,&=\,\, \frac{1}{2}\Delta(\overline B^0,\pi^-,D^+,\rho^0,\pi^-,D^+)
-\frac{1}{2}\Delta(\overline B^0,\pi^-,D^+,\omega,\pi^-,D^+),\\
L(E)[d,d,u,c]\,\,&=\,\, \frac{1}{2}\Delta(\overline B^0,\pi^-,D^+,\rho^0,\pi^-,D^+)
+\frac{1}{2}\Delta(\overline B^0,\pi^-,D^+,\omega,\pi^-,D^+).
\end{align}
Summing all of them, the rescattering amplitude in the $\overline B^0\to D^+\pi^-$ decay is
\begin{align}
\mathcal{A}_L(\overline B^0\to D^+\pi^-)=&L(T)_1[d,c,u,d]+L(E)[d,d,u,c]
=\Delta(\overline B^0,\pi^-,D^+,\rho^0,\pi^-,D^+).
\end{align}
The topological amplitude of $\overline B^0\to D^0\pi^0$ decay is $(E-C)/\sqrt{2}$.
The rescattering contributions include
\begin{align}
-\frac{1}{\sqrt{2}}L(C)_1[d,d,u,c]\,\,&=\,\, \frac{1}{\sqrt{2}}\Delta(\overline B^0,\pi^-,D^+,\rho^-,\pi^0,D^0),\\
-\frac{1}{\sqrt{2}}L(C)_2[d,d,u,c]\,\,&=\,\, \frac{1}{\sqrt{2}}\Delta(\overline B^0,\pi^-,D^+,D^{*-},D^0,\pi^0),\\
\frac{1}{\sqrt{2}}L(E)[d,u,u,c]\,\,&=\,\, \frac{1}{\sqrt{2}}\Delta(\overline B^0,\pi^-,D^+,\rho^-,\pi^0,D^0).
\end{align}
Summing all of them, the rescattering amplitude in the $\overline B^0\to D^0\pi^0$ decay is
\begin{align}
\mathcal{A}_L(\overline B^0\to D^0\pi^0)=&-\frac{1}{\sqrt{2}}L(C)_1[d,d,u,c]-\frac{1}{\sqrt{2}}L(C)_2[d,d,u,c]
+\frac{1}{\sqrt{2}}L(E)[d,u,u,c]\nonumber\\
=& \sqrt{2}\Delta(\overline B^0,\pi^-,D^+,\rho^-,\pi^0,D^0)+\frac{1}{\sqrt{2}}\Delta(\overline B^0,\pi^-,D^+,D^{*-},D^0,\pi^0).
\end{align}
The topological amplitude of $B^-\to D^0\pi^-$ decay is $T+C$.
The rescattering contributions include
\begin{align}
L(T)_2[u,c,u,d]\,\,&=\,\, -\frac{1}{2}\Delta(B^-,\pi^-, D^0,\rho^0,\pi^-, D^0)+\frac{1}{2}\Delta(B^-,\pi^-, D^0,\omega,\pi^-, D^0),\\
L(C)_1[u,d,u,c]\,\,&=\,\, -\frac{1}{2}\Delta(B^-,\pi^-, D^0,\rho^0,\pi^-, D^0)-\frac{1}{2}\Delta(B^-,\pi^-, D^0,\omega,\pi^-, D^0),\\
L(C)_2[u,d,u,c]\,\,&=\,\, -\Delta(B^-,\pi^-, D^0,D^{*-},D^0,\pi^-).
\end{align}
Summing all of them, the rescattering amplitude in the $B^-\to D^0\pi^-$ decay is
\begin{align}
\mathcal{A}_L(B^-\to D^0\pi^-)=&L(T)_2[u,c,u,d]+L(C)_1[u,d,u,c]+L(C)_2[u,d,u,c]\nonumber\\
=& -\Delta(B^-,\pi^-, D^0,\rho^0,\pi^-, D^0)-\Delta(B^-,\pi^-, D^0,D^{*-},D^0,\pi^-).
\end{align}

Under the Isospin symmetry, we have
\begin{align}
\Delta^\prime_1&= \Delta(\overline B^0,\pi^-,D^+,\rho^0,\pi^-,D^+) = \Delta(\overline B^0,\pi^-,D^+,\rho^-,\pi^0,D^0) = \Delta(B^-,\pi^-, D^0,\rho^0,\pi^-, D^0), \nonumber\\
  \Delta^\prime_2&=\Delta(\overline B^0,\pi^-,D^+,D^{*-},D^0,\pi^0) = \Delta(B^-,\pi^-, D^0,D^{*-},D^0,\pi^-).
\end{align}
The decay amplitudes of $\overline B^0\to D^+\pi^-$, $\overline B^0\to D^0\pi^0$ and $B^-\to D^0\pi^-$ channels can be written as
\begin{align}
\mathcal{A}_L(\overline B^0\to D^+\pi^-)& =\Delta^\prime_1,\qquad \mathcal{A}_L(\overline B^0\to D^0\pi^0)=\sqrt{2}\Delta^\prime_1+\frac{1}{\sqrt{2}}\Delta^\prime_2, \nonumber\\
\mathcal{A}_L(B^-\to D^0\pi^-)&=-\Delta^\prime_1-\Delta^\prime_2.
\end{align}
One can check the Isospin relation
\begin{align}
 \mathcal{A}(B^-\to D^0\pi^-)+\sqrt{2}\mathcal{A}(\overline B^0\to D^0\pi^0)=\mathcal{A}(\overline B^0\to D^+\pi^-)
\end{align}
is satisfied.
Under the $SU(4)_F$ symmetry, all the triangle diagrams are equivalent. The rescattering contributions of $T$ diagrams are zero, and $L(C)=-2\,L(E)$.

\subsection{$\overline B\to \pi\pi$ decay}\label{bpipi}

The topological amplitude of $\overline B^0\to \pi^+\pi^-$ decay is $\lambda_u(T+E)+\lambda_u(P+2PA)+\lambda_c(P+2PA)$, in which $\lambda_u=V_{ud}V^*_{ub}$ and $\lambda_c=V_{cd}V^*_{cb}$.
The rescattering contributions modeled by triangle diagram include
\begin{align}
\lambda_uL(T)_1[d,u,u,d]\,\,&=\,\, \frac{1}{2}\lambda_u\Delta(\overline B^0,\pi^-,\pi^+,\rho^0,\pi^-,\pi^+)
-\frac{1}{2}\lambda_u\Delta(\overline B^0,\pi^-,\pi^+,\omega,\pi^-,\pi^+),\\
\lambda_uL(T)_4[d,u,u,d]\,\,&=\,\, \frac{1}{2}\lambda_u\Delta(\overline B^0,\pi^-,\pi^+,\rho^0,\pi^-,\pi^+)
-\frac{1}{2}\lambda_u\Delta(\overline B^0,\pi^-,\pi^+,\omega,\pi^-,\pi^+),\\
\lambda_uL(E)[d,u,d,u]\,\,&=\,\, \frac{1}{2}\lambda_u\Delta(\overline B^0,\pi^-,\pi^+,\rho^0,\pi^-,\pi^+)
+\frac{1}{2}\lambda_u\Delta(\overline B^0,\pi^-,\pi^+,\omega,\pi^-,\pi^+),\\
\lambda_uL(P)[d,u,d,u]\,\,&=\,\, \frac{1}{2}\lambda_u\Delta(\overline B^0,\pi^-,\pi^+,\rho^0,\pi^-,\pi^+)
+\frac{1}{2}\lambda_u\Delta(\overline B^0,\pi^-,\pi^+,\omega,\pi^-,\pi^+),\\
\lambda_cL(P)[d,u,d,c]\,\,&=\,\, \lambda_c\Delta(\overline B^0,D^-,D^+,\overline D^{*0},\pi^-,\pi^+).
\end{align}
Summing all of them, the rescattering amplitude in the $\overline B^0\to \pi^+\pi^-$ decay is
\begin{align}
\mathcal{A}_L(\overline B^0\to \pi^+\pi^-)=&\lambda_u(L(T)_1[d,u,u,d]+L(T)_4[d,u,u,d]+L(E)[d,u,d,u]\nonumber\\
&+L(P)[d,u,d,u])+\lambda_cL(P)[d,u,d,c]\nonumber\\
=& 2\lambda_u\Delta(\overline B^0,\pi^-,\pi^+,\rho^0,\pi^-,\pi^+)+\lambda_c\Delta(\overline B^0,D^-,D^+,\overline D^{*0},\pi^-,\pi^+).
\end{align}
The topological amplitude of $\overline B^0\to \pi^0\pi^0$ decay is $\lambda_u(E-C)/\sqrt{2}+\lambda_u(P+2PA)/\sqrt{2}+\lambda_c(P+2PA)/\sqrt{2}$.
The rescattering contributions include
\begin{align}
\frac{1}{\sqrt{2}}\lambda_uL(E)[d,u,u,u]\,\,&=\,\, \frac{1}{\sqrt{2}}\lambda_u\Delta(\overline B^0,\pi^-,\pi^+,\rho^-,\pi^0,\pi^0),\\
-\frac{1}{\sqrt{2}}\lambda_uL(C)_1[d,d,u,u]\,\,&=\,\, \frac{1}{\sqrt{2}}\lambda_u\Delta(\overline B^0,\pi^-,\pi^+,\rho^-,\pi^0,\pi^0),\\
-\frac{1}{\sqrt{2}}\lambda_uL(C)_2[d,d,u,u]\,\,&=\,\, \frac{1}{\sqrt{2}}\lambda_u\Delta(\overline B^0,\pi^-,\pi^+,\rho^-,\pi^0,\pi^0),\\
\frac{1}{\sqrt{2}}\lambda_uL(P)[d,d,d,u]\,\,&=\,\, \frac{1}{\sqrt{2}}\lambda_u\Delta(\overline B^0,\pi^-,\pi^+,\rho^-,\pi^0,\pi^0),\\
\frac{1}{\sqrt{2}}\lambda_cL(P)[d,d,d,c]\,\,&=\,\, \frac{1}{\sqrt{2}}\lambda_c\Delta(\overline B^0,D^-,D^+,D^{*-},\pi^0,\pi^0).
\end{align}
Summing all of them, the rescattering amplitude in the $\overline B^0\to \pi^0\pi^0$ decay is
\begin{align}
\mathcal{A}_L(\overline B^0\to \pi^0\pi^0)=&\frac{1}{\sqrt{2}}\lambda_d(L(E)[d,u,u,u]-L(C)_1[d,d,u,u]
-L(C)_2[d,d,u,u]\nonumber\\&+L(P)[d,d,d,u])
+\frac{1}{\sqrt{2}}\lambda_cL(P)[d,d,d,c]\nonumber\\
=& 2\sqrt{2}\lambda_u\Delta(\overline B^0,\pi^-,\pi^+,\rho^-,\pi^0,\pi^0)+\frac{1}{\sqrt{2}}\lambda_c\Delta(\overline B^0,D^-,D^+,D^{*-},\pi^0,\pi^0).
\end{align}
The topological amplitude of $B^-\to \pi^-\pi^0$ decay is $\lambda_u(T+C)/\sqrt{2}$.
The rescattering contributions include
\begin{align}
\frac{1}{\sqrt{2}}\lambda_uL(T)_2[u,u,d,u]\,\,=&\,\, -\frac{1}{4\sqrt{2}}\lambda_u\Delta(B^-,\pi^-,\pi^0,\rho^0,\pi^-,\pi^0)
-\frac{1}{12\sqrt{2}}\lambda_u\Delta(B^-,\pi^-,\eta_8,\rho^0,\pi^-,\pi^0)\nonumber\\ & -\frac{1}{6\sqrt{2}}\lambda_u\Delta(B^-,\pi^-,\eta_1,\rho^0,\pi^-,\pi^0)
+\frac{1}{4\sqrt{2}}\lambda_u\Delta(B^-,\pi^-,\pi^0,\omega,\pi^-,\pi^0)\nonumber\\ & +\frac{1}{12\sqrt{2}}\lambda_u\Delta(B^-,\pi^-,\eta_8,\omega,\pi^-,\pi^0)
+\frac{1}{6\sqrt{2}}\lambda_u\Delta(B^-,\pi^-,\eta_1,\omega,\pi^-,\pi^0),\\
\frac{1}{\sqrt{2}}\lambda_uL(T)_4[u,u,d,u]\,\,=&\,\, \frac{1}{4\sqrt{2}}\lambda_u\Delta(B^-,\pi^-,\pi^0,\rho^0,\pi^-,\pi^0)
+\frac{1}{12\sqrt{2}}\lambda_u\Delta(B^-,\pi^-,\eta_8,\rho^0,\pi^-,\pi^0)\nonumber\\ & +\frac{1}{6\sqrt{2}}\lambda_u\Delta(B^-,\pi^-,\eta_1,\rho^0,\pi^-,\pi^0)
-\frac{1}{4\sqrt{2}}\lambda_u\Delta(B^-,\pi^-,\pi^0,\omega,\pi^-,\pi^0)\nonumber\\ & -\frac{1}{12\sqrt{2}}\lambda_u\Delta(B^-,\pi^-,\eta_8,\omega,\pi^-,\pi^0)
-\frac{1}{6\sqrt{2}}\lambda_u\Delta(B^-,\pi^-,\eta_1,\omega,\pi^-,\pi^0),
\end{align}
\begin{align}
\frac{1}{\sqrt{2}}\lambda_uL(C)_1[u,d,u,u]\,\,=&\,\, -\frac{1}{4\sqrt{2}}\lambda_u\Delta(B^-,\pi^-,\pi^0,\rho^0,\pi^-,\pi^0)
-\frac{1}{12\sqrt{2}}\lambda_u\Delta(B^-,\pi^-,\eta_8,\rho^0,\pi^-,\pi^0)\nonumber\\ & -\frac{1}{6\sqrt{2}}\lambda_u\Delta(B^-,\pi^-,\eta_1,\rho^0,\pi^-,\pi^0)
-\frac{1}{4\sqrt{2}}\lambda_u\Delta(B^-,\pi^-,\pi^0,\omega,\pi^-,\pi^0)\nonumber\\ & -\frac{1}{12\sqrt{2}}\lambda_u\Delta(B^-,\pi^-,\eta_8,\omega,\pi^-,\pi^0)
-\frac{1}{6\sqrt{2}}\lambda_u\Delta(B^-,\pi^-,\eta_1,\omega,\pi^-,\pi^0),\\
\frac{1}{\sqrt{2}}\lambda_uL(C)_2[u,d,u,u]\,\,=&\,\, -\frac{1}{2\sqrt{2}}\lambda_u\Delta(B^-,\pi^-,\pi^0,\rho^-,\pi^0,\pi^-)
-\frac{1}{6\sqrt{2}}\lambda_u\Delta(B^-,\pi^-,\eta_8,\rho^-,\pi^0,\pi^-)\nonumber\\ & -\frac{1}{3\sqrt{2}}\lambda_u\Delta(B^-,\pi^-,\eta_1,\rho^-,\pi^0,\pi^-),
\end{align}
\begin{align}
-\frac{1}{\sqrt{2}}\lambda_uL(A)_1[u,u,d,d]\,\,=&\,\, -\frac{1}{4\sqrt{2}}\lambda_u\Delta(B^-,\pi^-,\pi^0,\rho^0,\pi^-,\pi^0)
+\frac{1}{12\sqrt{2}}\lambda_u\Delta(B^-,\pi^-,\eta_8,\rho^0,\pi^-,\pi^0)\nonumber\\ & +\frac{1}{6\sqrt{2}}\lambda_u\Delta(B^-,\pi^-,\eta_1,\rho^0,\pi^-,\pi^0)
-\frac{1}{4\sqrt{2}}\lambda_u\Delta(B^-,\pi^-,\pi^0,\omega,\pi^-,\pi^0)\nonumber\\ & +\frac{1}{12\sqrt{2}}\lambda_u\Delta(B^-,\pi^-,\eta_8,\omega,\pi^-,\pi^0)
+\frac{1}{6\sqrt{2}}\lambda_u\Delta(B^-,\pi^-,\eta_1,\rho^0,\pi^-,\pi^0),\\
-\frac{1}{\sqrt{2}}\lambda_uL(A)_2[u,u,d,d]\,\,=&\,\, -\frac{1}{2\sqrt{2}}\lambda_u\Delta(B^-,\pi^-,\pi^0,\rho^-,\pi^0,\pi^-)
+\frac{1}{6\sqrt{2}}\lambda_u\Delta(B^-,\pi^-,\eta_8,\rho^-,\pi^0,\pi^-)\nonumber\\ & +\frac{1}{3\sqrt{2}}\lambda_u\Delta(B^-,\pi^-,\eta_1,\rho^-,\pi^0,\pi^-),\\
-\frac{1}{\sqrt{2}}\lambda_uL(A)_3[u,u,u,d]\,\,=&\,\, +\frac{1}{4\sqrt{2}}\lambda_u\Delta(B^-,\pi^-,\pi^0,\rho^0,\pi^-,\pi^0)
-\frac{1}{12\sqrt{2}}\lambda_u\Delta(B^-,\pi^-,\eta_8,\rho^0,\pi^-,\pi^0)\nonumber\\ & -\frac{1}{6\sqrt{2}}\lambda_u\Delta(B^-,\pi^-,\eta_1,\rho^0,\pi^-,\pi^0)
+\frac{1}{4\sqrt{2}}\lambda_u\Delta(B^-,\pi^-,\pi^0,\omega,\pi^-,\pi^0)\nonumber\\ & -\frac{1}{12\sqrt{2}}\lambda_u\Delta(B^-,\pi^-,\eta_8,\omega,\pi^-,\pi^0)
-\frac{1}{6\sqrt{2}}\lambda_u\Delta(B^-,\pi^-,\eta_1,\rho^0,\pi^-,\pi^0),\\
\frac{1}{\sqrt{2}}\lambda_uL(A)_3[u,u,d,d]\,\,=&\,\, -\frac{1}{2\sqrt{2}}\lambda_u\Delta(B^-,\pi^-,\pi^0,\rho^-,\pi^0,\pi^-)
+\frac{1}{6\sqrt{2}}\lambda_u\Delta(B^-,\pi^-,\eta_8,\rho^-,\pi^0,\pi^-)\nonumber\\ & +\frac{1}{3\sqrt{2}}\lambda_u\Delta(B^-,\pi^-,\eta_1,\rho^-,\pi^0,\pi^-)),
\end{align}
\begin{align}
\frac{1}{\sqrt{2}}\lambda_uL(P)[u,u,d,u]\,\,=&\,\, \frac{1}{4\sqrt{2}}\lambda_u\Delta(B^-,\pi^-,\pi^0,\rho^0,\pi^-,\pi^0)
+\frac{1}{12\sqrt{2}}\lambda_u\Delta(B^-,\pi^-,\eta_8,\rho^0,\pi^-,\pi^0)\nonumber\\ & +\frac{1}{6\sqrt{2}}\lambda_u\Delta(B^-,\pi^-,\eta_1,\rho^0,\pi^-,\pi^0)
+\frac{1}{4\sqrt{2}}\lambda_u\Delta(B^-,\pi^-,\pi^0,\omega,\pi^-,\pi^0)\nonumber\\ & +\frac{1}{12\sqrt{2}}\lambda_u\Delta(B^-,\pi^-,\eta_8,\omega,\pi^-,\pi^0)
+\frac{1}{6\sqrt{2}}\lambda_u\Delta(B^-,\pi^-,\eta_1,\rho^0,\pi^-,\pi^0),\\
-\frac{1}{\sqrt{2}}\lambda_uL(P)[u,d,d,u]\,\,=&\,\, -\frac{1}{2\sqrt{2}}\lambda_u\Delta(B^-,\pi^-,\pi^0,\rho^-,\pi^0,\pi^-)
-\frac{1}{6\sqrt{2}}\lambda_u\Delta(B^-,\pi^-,\eta_8,\rho^-,\pi^0,\pi^-)\nonumber\\ & -\frac{1}{3\sqrt{2}}\lambda_u\Delta(B^-,\pi^-,\eta_1,\rho^-,\pi^0,\pi^-),\\
\frac{1}{\sqrt{2}}\lambda_cL(P)[u,u,d,c]\,\,&=\,\, \frac{1}{\sqrt{2}}\lambda_c\Delta(B^-,D^-,D^0,\overline D^{*0},\pi^-,\pi^0),\\
-\frac{1}{\sqrt{2}}\lambda_cL(P)[u,d,d,c]\,\,&=\,\, -\frac{1}{\sqrt{2}}\lambda_c\Delta(B^-,D^-,D^0,D^{*-},\pi^0,\pi^-).
\end{align}
Summing all topological-scattering diagrams, the rescattering amplitude in the $B^-\to \pi^-\pi^0$ decay is
\begin{align}
\mathcal{A}_L(B^-\to \pi^-\pi^0)=&\frac{1}{\sqrt{2}}\lambda_u(L(T)_2[u,u,d,u]+L(T)_4[u,u,d,u]+L(C)_1[u,d,u,u]\nonumber\\&
+L(C)_2[u,d,u,u]-L(A)_1[u,u,d,d]-L(A)_2[u,u,d,d]\nonumber\\&-L(A)_3[u,u,u,d]+L(A)_3[u,u,d,d]
+L(P)[u,u,d,u]\nonumber\\&
-L(P)[u,d,d,u])+\frac{1}{\sqrt{2}}\lambda_c(L(P)[u,u,d,c]-L(P)[u,d,d,c])\nonumber\\=&
 -\sqrt{2}\lambda_u\Delta(B^-,\pi^-,\pi^0,\rho^-,\pi^0,\pi^-)
 +\frac{1}{\sqrt{2}}\lambda_c(\Delta(B^-,D^-,D^0,\overline D^{*0},\pi^-,\pi^0)\nonumber\\&
-\Delta(B^-,D^-,D^0,D^{*-},\pi^0,\pi^-)).
\end{align}

Under the Isospin symmetry, we have
\begin{align}
\Delta^\prime_3&= \Delta(\overline B^0,\pi^-,\pi^+,\rho^0,\pi^-,\pi^+) = \Delta(\overline B^0,\pi^-,\pi^+,\rho^-,\pi^0,\pi^0) = \Delta(B^-,\pi^-,\pi^0,\rho^-,\pi^0,\pi^-), \nonumber\\
  \Delta^\prime_4&=\Delta(\overline B^0,D^-,D^+,\overline D^{*0},\pi^-,\pi^+) = \Delta(\overline B^0,D^-,D^+,D^{*-},\pi^0,\pi^0)\nonumber\\&~~~~~~=\Delta(B^-,D^-,D^0,\overline D^{*0},\pi^-,\pi^0)=\Delta(B^-,D^-,D^0,D^{*-},\pi^0,\pi^-).
\end{align}
The decay amplitudes of $\overline B^0\to \pi^+\pi^-$, $\overline B^0\to \pi^0\pi^0$ and $B^-\to \pi^-\pi^0$ channels can be written as
\begin{align}
\mathcal{A}_L(\overline B^0\to \pi^+\pi^-)& =2\lambda_u\Delta^\prime_3+\lambda_c\Delta^\prime_4,\qquad \mathcal{A}_L(\overline B^0\to \pi^0\pi^0)=2\sqrt{2}\lambda_u\Delta^\prime_3+\frac{1}{\sqrt{2}}\lambda_c\Delta^\prime_4, \nonumber\\
\mathcal{A}_L(B^-\to \pi^-\pi^0)&=-\sqrt{2}\lambda_u\Delta^\prime_3.
\end{align}
One can check the Isospin relation is maintained
\begin{align}
 \mathcal{A}(\overline B^0\to \pi^+\pi^-) = \sqrt{2}\mathcal{A}(\overline B^0\to \pi^0\pi^0)+\sqrt{2}\mathcal{A}(B^-\to \pi^-\pi^0).
\end{align}

\end{appendix}

\end{document}